\documentclass[pre,nofootinbib,superscriptaddress,twocolumn,floatfix,longbibliography]{revtex4-1}
\usepackage{graphicx}
\usepackage{amsmath}
\usepackage{amssymb}
\usepackage{tikz}
\usepackage{color} 
\usepackage{comment}
\usepackage{caption}
\usepackage{subcaption}
\usepackage{textcomp}

\newcommand{\rs}[1]{\textcolor{black}{#1}}
\usepackage{hyperref}
\hypersetup{colorlinks=true,linkcolor=blue,urlcolor=blue}

\begin{document}

\title{Breaking Mechanical Holography in Combinatorial Metamaterials}

\author{Chaviva Sirote-Katz}
\email{chavivas@mail.tau.ac.il}
\affiliation{Department of Biomedical Engineering, Tel Aviv University, Tel Aviv 69978, Israel}

\author{Ofri Palti}
\affiliation{The Future Scientists Center–Alpha Program, Tel Aviv Youth University, Tel Aviv 69978, Israel}

\author{Naomi Spiro}
\affiliation{University of Illinois at Urbana-Champaign, Urbana, IL 61801, USA}

\author{Tam\'as K\'alm\'an}
\email{kalman@math.titech.ac.jp}
\affiliation{Department of Mathematics, Institute of Science Tokyo, H-214, 2-12-1 Ookayama, Meguro-ku, Tokyo 152-8551, Japan}
\affiliation{International Institute for Sustainability with Knotted Chiral Meta Matter (WPI-SKCM$^2$), Hiroshima University, Higashi-Hiroshima, Hiroshima 739-8526, Japan}

\author{Yair Shokef}
\email{shokef@tau.ac.il}
\affiliation{School of Mechanical Engineering, Tel Aviv University, Tel Aviv 69978, Israel}
\affiliation{School of Physics and Astronomy, Tel Aviv University, Tel Aviv 69978, Israel}
\affiliation{Center for Computational Molecular and Materials Science, Tel Aviv University, Tel Aviv 69978, Israel}
\affiliation{Center for Physics and Chemistry of Living Systems, Tel Aviv University, 69978, Tel Aviv, Israel}
\affiliation{International Institute for Sustainability with Knotted Chiral Meta Matter (WPI-SKCM$^2$), Hiroshima University, Higashi-Hiroshima, Hiroshima 739-8526, Japan}

\begin{abstract}

Combinatorial mechanical metamaterials are made of anisotropic, flexible blocks, such that multiple metamaterials may be constructed using a single block type, and the system's response strongly depends on the mutual orientations of the blocks within the lattice. We study a family of possible block types for the square, honeycomb, and cubic lattices. Blocks that are centrally symmetric induce holographic order, such that mechanical compatibility (meaning that blocks do not impede each other's motion) implies bulk-boundary coupling. With them, one can design a compatible metamaterial that will deform in any desired texture only on part of its boundary. With blocks that break holographic order, we demonstrate how to design the deformation texture on the entire boundary. Correspondingly, the number of compatible holographic metamaterials scales exponentially with the boundary, while in non-holographic cases we show that it scales exponentially with the bulk. 

\end{abstract}

\maketitle

\section{Introduction} 
\label{sec:intro}

In mechanical metamaterials we study structures that repeat in a lattice, such that the geometric deformation of each building block and the resultant interaction between neighboring blocks lead to non-trivial cooperative mechanical responses at the system level~\cite{bertoldi_flexible_review_2017, jang_soft_skin_2015, czajkowski_conformal_2022, guo_non-orientable_2023, chaco, kwakernaak_counting_2023}. Conventionally, and inspired by natural crystalline materials, most metamaterials are spatially periodic. However, since metamaterials are artificially designed and then fabricated at the macroscopic scale, one can construct not only any crystal- or quasi-crystal structure, but also arbitrary non-periodic structures. An approach to systematically consider such complex structures is based on anisotropic building blocks positioned on a regular lattice, where each repeating block may be independently oriented along any one of the principal directions of the lattice. The resulting class of metamaterials is referred to as \emph{combinatorial metamaterials}~\cite{coulais2016metacube}, since the possible structures are defined by the discrete orientations of all blocks in the lattice, with the total number of possible metamaterials growing exponentially with the number of blocks in the lattice.

In combinatorial metamaterials, each building block typically has one soft mode of deformation and the mutual orientations of neighboring blocks determine if they can all simultaneously deform according to their soft modes, thus forming a mechanically \emph{compatible} structure with a spatially extended soft mode of deformation. Alternatively, if neighboring blocks frustrate each other's ability to easily deform, the metamaterial lacks such a global soft mode and is thus more rigid and harder to deform. Beyond the potential applications of such mechanical systems, their study is useful for understanding the effects of frustration and how to avoid it in other physical systems~\cite{blunt2008, Morrison2013, Kang2014, oguz2020, pisanty2021putting, hanai2024}. Mechanical compatibility leads to combinatorial matching rules between neighboring units, with similar concepts employed in magnetism~\cite{villain1977} and in origami~\cite{dieleman_jigsaw_2020}. Manipulating the discrete orientations of the building blocks allows to design metamaterials with textured response~\cite{coulais2016metacube} or to use topological defects for spatially steering stress fields~\cite{meeussen2020supertriangles, pisanty2021putting}. Computational~\cite{bossart_oligomodal_2021, van_mastrigt_machine_2022, van_mastrigt_emergent_2023} \rs{and theoretical}~\cite{T1T2} methods have been employed to obtain multiple desired responses of a given metamaterial when there are two soft modes per building block.

Previous works considered cubic blocks forming a three-dimensional (3D) cubic lattice~\cite{coulais2016metacube}, triangular blocks forming a two-dimensional (2D) triangular lattice~\cite{meeussen2020supertriangles, meeussen_NJP_2020, T1T2}, hexagonal blocks forming a 2D honeycomb lattice~\cite{pisanty2021putting}, and square blocks forming a 2D square lattice~\cite{bossart_oligomodal_2021, van_mastrigt_machine_2022, van_mastrigt_emergent_2023}. For the cubic and the hexagonal blocks studied so far, each block's soft mode of deformation had each pair of opposing facets move either toward or away from each other. Hence, when these blocks were combined together in a compatible manner, the global soft mode of the resulting metamaterial had alternating displacements along any line running through the lattice in any one of its principal directions. In particular, the texture of deformation on the boundary of the metamaterial determined the deformations within the bulk of the system, thus constituting the \emph{holographic order} that such systems exhibit. In addition to this alternating behavior, we will also see examples of \emph{persistent holography}, where displacements along all principal directions are uniformly parallel to each other. Then, there are yet other block types that exhibit neither type of holographic order.

One might naturally wonder about the number of compatible metamaterials, hereafter referred to as the \emph{multiplicity}, which a given region can support. Since each deformation field corresponds to a unique orientation of all blocks, in holographic systems, the multiplicity of compatible metamaterials scales exponentially with the boundary (perimeter in 2D or surface area in 3D) of the lattice, rather than with the number of blocks within its bulk (i.e., the area in 2D or volume in 3D). Triangular blocks do not induce holographic order, since a triangle has only one side along each one of the triangular lattice's principal directions. Interestingly, compatible combinatorial metamaterials constructed from these triangles were shown to have multiplicity that grows exponentially with the total number of blocks in the lattice~\cite{meeussen2020supertriangles}. However, a comprehensive connection between holography and the multiplicity of compatible metamaterials has not yet been established.

In this paper we present a unified framework for generating both holographic and non-holographic combinatorial metamaterials in 2D and in 3D, by considering all possible building blocks for the 2D square and honeycomb lattices, as well as for the 3D cubic lattice. In Sec.~\ref{sec:blocks} we systematically identify all possible blocks and we analyze their deformations along the different principal directions to recognize blocks that will induce holographic order vs.\ those that will not. Before delving into the emergent consequences of constructing metamaterials from blocks with different mechanical responses, in Sec.~\ref{sec:realization} we complement our theoretical analysis by presenting, for the three lattices that we study, actual mechanical designs and experimental prototypes for the internal structures that give the desired mechanical functionalities of the different block types. Then, in Sec.~\ref{sec:compatibility} we theoretically identify the conditions for mechanical compatibility. In Sec.~\ref{sec:multiplicity} we analyze the multiplicity of compatible metamaterials, and show that in the presence of holographic order, it scales exponentially with the boundary of the system, while for all non-holographic cases considered here, it scales exponentially with the total number of blocks in the system. In Sec.~\ref{sec:texture} we design non-holographic metamaterials that can deform to arbitrary textures on their entire boundary. This is in contrast to previous work on the cubic lattice with holographic order, where the texture only on one face could be freely designed.  Finally, in Sec.~\ref{sec:discussion} we discuss the results, and their consequences for further work, on characterizing and designing physical responses of combinatorial metamaterials.

Throughout this paper we focus on compatible metamaterials, in which the blocks are positioned and oriented such that there would not be any frustration in the system. In an accompanying paper~\cite{defects_paper} we study frustrated metamaterials that may be constructed using the blocks introduced here, and we ask, both in 2D and in 3D, whether mechanical defects may be positioned in any desired spatial distribution. We show that this is possible only for some of the block types, while for others it is not the case. In 3D, we show that even when not any geometry of defects may be realized, the blocks may be oriented so as to construct any topology (meaning knot or link type) for the defect lines.

\section{Building Blocks} 
\label{sec:blocks}

A given lattice structure implies a certain shape for its building blocks. Here, we restrict ourselves to blocks with a single soft mode, in which the motion of each of the block’s facets is either into or out of the block; all motions are orthogonal to the facet; and all have the same magnitude. Even with these restrictions, the soft mode of deformation of the block can take different discrete forms, which are characterized by whether each facet moves in or out with respect to the block. In this section we introduce a common framework for describing all possible blocks in different lattices, and specifically focus on the 2D square lattice, the 3D cubic lattice and the 2D honeycomb lattice.

\subsection{Square Lattice}

\begin{figure}[t!]
\centering
\includegraphics[width=\columnwidth]{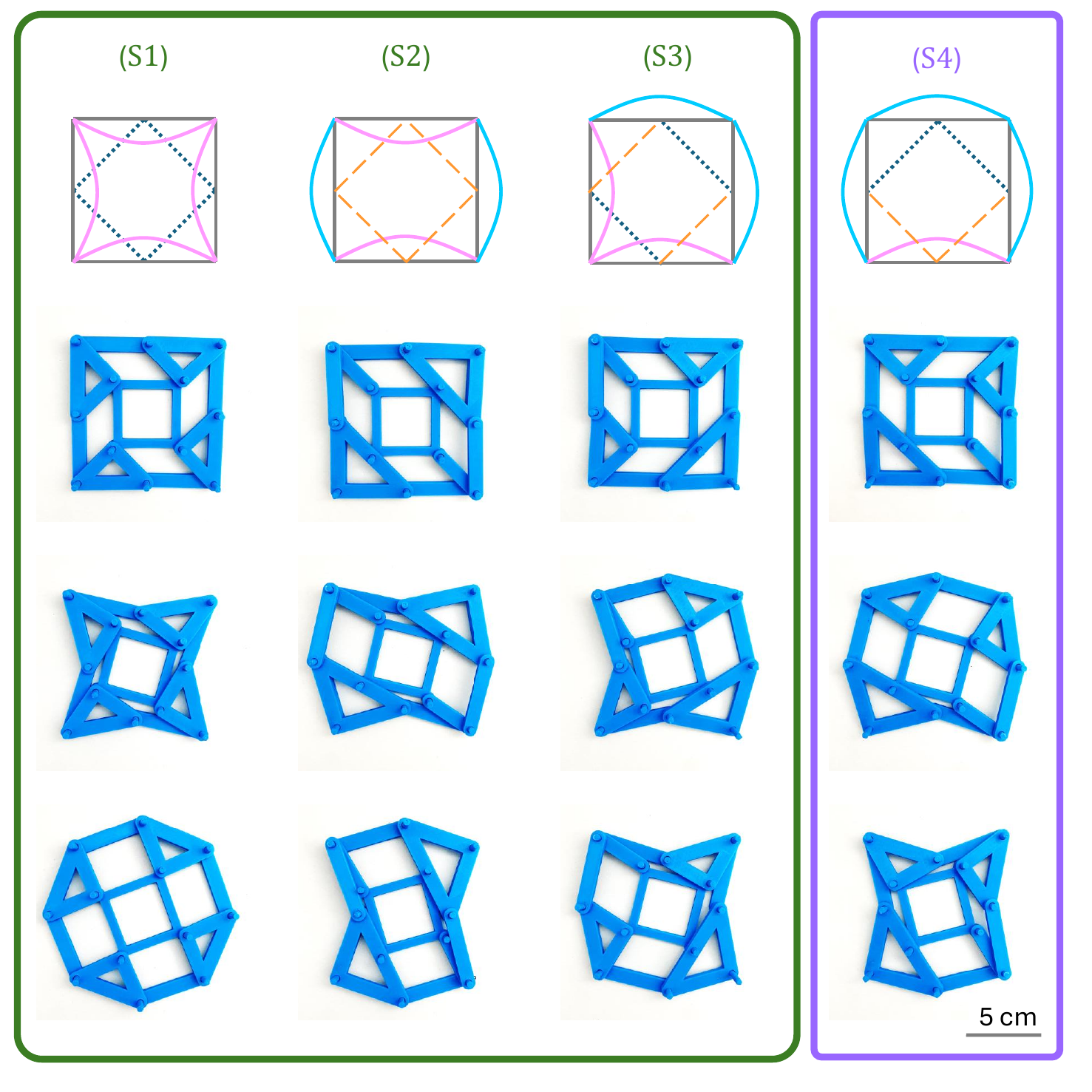}
\caption{\textbf{Square blocks.} For each of the four possible blocks, we draw the struts (orange) and hinges (blue) that couple neighboring sides, and mark one of the two polarizations of the soft mode of deformation (top), suggest a mechanical realization (center), and show its two deformations (bottom). Green and purple frames indicate holographic and non-holographic blocks, respectively.}
\label{fig:square_blocks}
\end{figure}

Perhaps the most widely studied mechanical metamaterial is based on the 2D square lattice, with a repeating structure that deforms most easily in the quadrupolar manner shown as Block~S2 in Fig.~\ref{fig:square_blocks}, see~\cite{resch_1965_patent, bertoldi_AdvMat2010, bar-sinai_charges_2020, Deng_domain_walls_2020, merrigan_PRR2021, czajkowski_conformal_2022}. We extend this to consider all possible mechanical functionalities of a square building block. For a square block, each side can go in or out, leading to a total of $2^4=16$ possible deformations. However, deformation patterns that are related by inversion of all displacements correspond to the two possible \emph{polarizations} of a single block. In other words, these are the two directions in which the soft mode of the block may be actuated. Moreover, by rotational symmetry, the possible blocks reduce to the four shown in Fig.~\ref{fig:square_blocks}. 

In the drawings of Fig.~\ref{fig:square_blocks}, in addition to schematically marking the two possible deformation directions of each side by a pink or blue arc, we draw a dashed orange line between adjacent sides such that when one contracts in to the block, the other protrudes out of it. This pair of sides may be thought of as being connected with a \emph{strut} that forces the two to move together. Conversely, when a pair of adjacent sides tends to both move in to the block or both out of it, we connect them with a dotted blue line, which represents a \emph{hinge}, where there is preference for bending the angle between the two sides, either to contract it or to expand it. To simplify subsequent figures, later on we will omit the strut lines and will mark only the hinge lines. We can use the struts and hinges to recognize that Blocks~S1 and~S2 are isotropic, while Block~S3 has two possible orientations and Block~S4 has four possible orientations. 

The block types may be classified according to the deformations along the principal axes of the building block. For Blocks~S1 and~S2, opposite sides move in opposite directions. Thus, in a compatible metamaterial made of Block~S1, as well as in one made of Block~S2, displacements will perfectly alternate along any row or column in the lattice. Similarly, for Block~S3, opposite sides move in the same direction. Thus, in a compatible metamaterial made of Block~S3, even though it is anisotropic, displacements along any row or column in the lattice will all be in the same direction, albeit possibly in different directions in different rows and columns. Therefore, square Blocks~S1-S3 exhibit holographic order, since the direction of deformation on the surface of the metamaterial made of these blocks fully determines the deformations in the bulk of the metamaterial. This is indicated by the green frame around them in Fig.~\ref{fig:square_blocks}.

We will refer to holographic order in blocks like~S1 and~S2 as \emph{alternating}, because the displacement direction alternates as one moves through the lattice. Meanwhile, holographic order as in Block~S3 will be called \emph{persistent}, since in this case the direction of displacement within the metamaterial persists as one moves along any line in the lattice.

In contrast to these holographic blocks, Block~S4 has displacements in the same direction at one pair of its opposing sides, and displacements in opposite directions at the other pair, in other words it imposes persistence in one direction and alternation in the other direction. Since each block within the lattice can be oriented independently, in a compatible metamaterial made of these blocks, the displacements on the surface of the lattice do not determine the displacements in the bulk, that is, such a metamaterial is non-holographic, and in Fig.~\ref{fig:square_blocks} we encircle it with a purple frame.

\subsection{Cubic Lattice}

\begin{figure*}[t!]
\centering
\includegraphics[width=\textwidth]{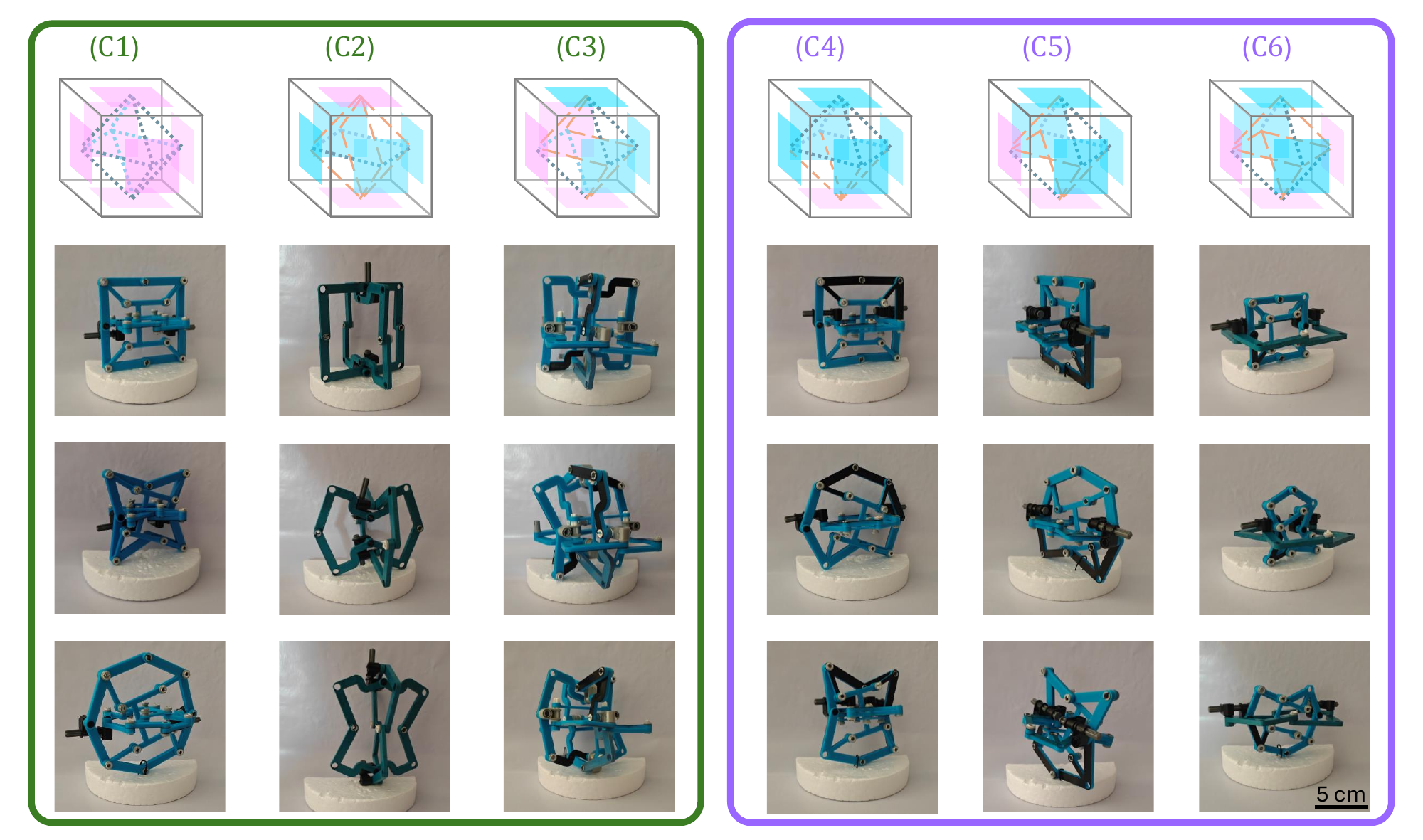}
\caption{\textbf{Cubic blocks.} For each of the six possible blocks, we draw the struts (orange) and hinges (blue) that couple neighboring sides, and mark one of the two polarizations of the soft mode of deformation by pink and blue colors for the two possible directions of motion of each face (top). We suggest a mechanical realization (bottom). Green and purple frames indicate holographic and non-holographic blocks, respectively.}
\label{fig:cubic_blocks}
\end{figure*}

We extend the above to 3D by considering the simple cubic lattice. As before, we assume the same magnitude of deformation along all six faces of a cubic building block, and use rotational symmetry to reduce the total number of $2^6 = 64$ displacement states to the six blocks shown in Fig.~\ref{fig:cubic_blocks}. Here as well, we denote by dashed orange lines the struts that connect adjacent faces so that when one moves into the block the other moves out of it; and we denote by dotted blue lines the hinges that connect adjacent faces such that both move in to the block or both move out of it. We see that Block~C1 is isotropic, Block~C2 has three possible orientations, Block~C3 has four orientations, Blocks~C4 and C6 each have six possible orientations, and Block~C5 is the least symmetric, with $12$ possible orientations.

Similarly to the square lattice, we identify all possible blocks by considering the displacements in the three principal directions. For blocks that exhibit holographic order, there is full correspondence between Blocks~S1-S3 in the square lattice and Blocks~C1-C3 in the cubic lattice: Blocks~S1 and~C1 induce alternating holographic order, and the motion is in opposite directions for all pairs of opposite facets, namely all in or all out. For Blocks~S2 and~C2, the motion is still in opposite directions for each axis, but one of the axes behaves differently from the other(s). Despite the fact that there are more axes in the cubic lattice, there are no further possibilities here. For Blocks~S3 and~C3, which induce persistent holographic order, the motion is along the same direction along all axes, and here too, there is no further richness in the 3D cube compared to the 2D square. However, in the cubic lattice, there are more non-holographic blocks than the single such block in the square lattice. For the 2D square, holography is broken by having persistent displacements for one axis, and alternating motion for the other axis, as exhibited by Block~S4. For the 3D cube, 
we can have one axis of one type and two of the other type, furthermore when there are two alternating axes, they may or may not be synchronized with each other. This results in three possible ways of non-holographic behavior, as seen in Blocks~C4-C6. Similarly to the square blocks, this classification for the cubic blocks is indicated in Fig.~\ref{fig:cubic_blocks} by the green frame around the holographic blocks and the purple frame around the non-holographics ones.

One can go beyond just the analogy between the 2D and 3D cases by noting that each cubic block is made of three orthogonal squares, and the behavior in each of these squares matches one of the square blocks defined above. Specifically, we have $\mathrm C1=(\mathrm S1,\mathrm S1,\mathrm S1)$, $\mathrm C2=(\mathrm S1,\mathrm S2,\mathrm S2)$, $\mathrm C3=(\mathrm S3,\mathrm S3,\mathrm S3)$, $\mathrm C4=(\mathrm S1,\mathrm S4,\mathrm S4)$, $\mathrm C5=(\mathrm S4,\mathrm S4,\mathrm S3)$, and $\mathrm C6=(\mathrm S2,\mathrm S4,\mathrm S4)$ in this sense. We will utilize this idea when designing the texture on the surface of a 3D metamaterial, and when physically constructing the 3D blocks by combining 2D blocks in orthogonal planes.

\subsection{Honeycomb Lattice}

\begin{figure*}[t!]
\centering
\includegraphics[width=\textwidth]{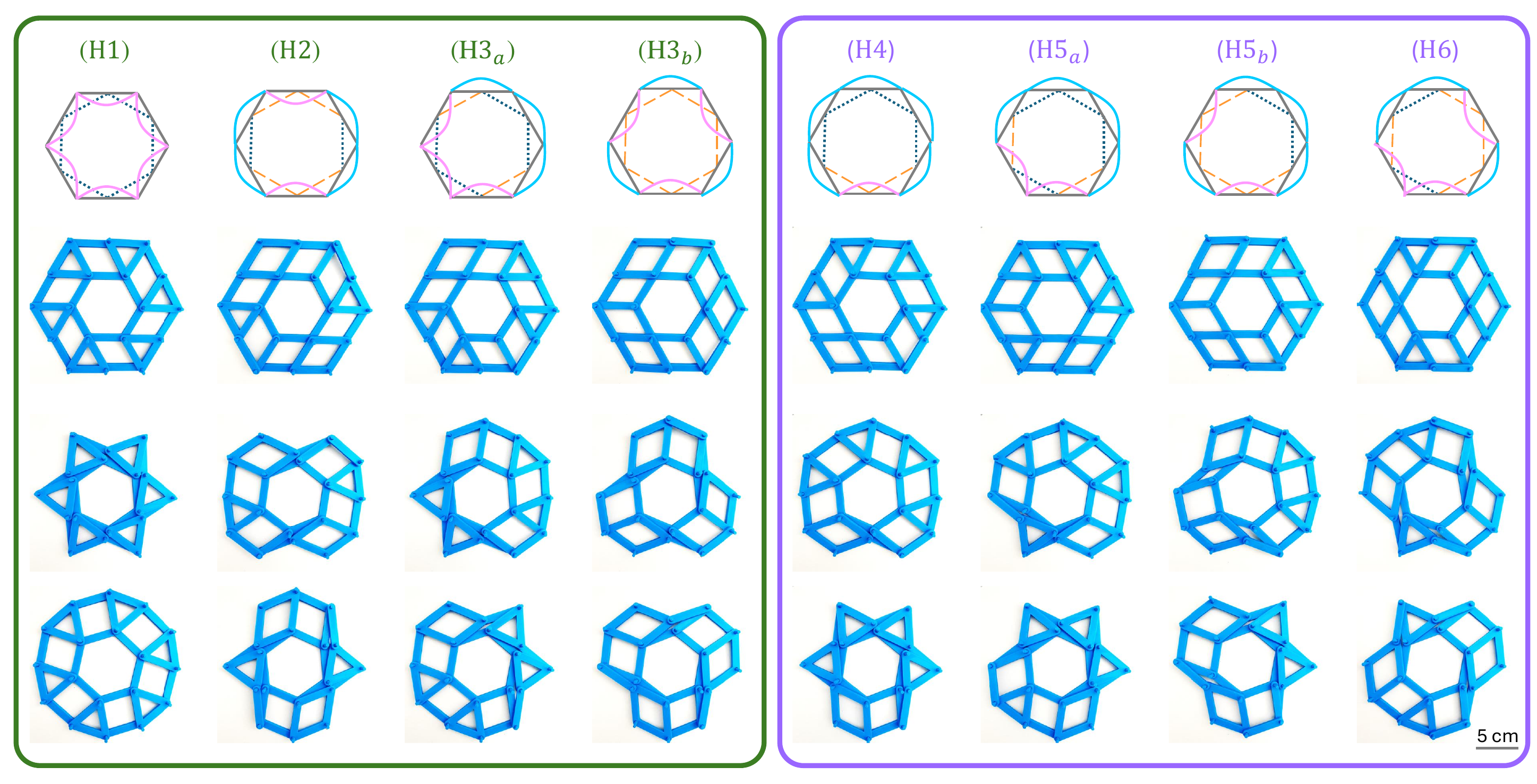}
\caption{\textbf{Hexagonal blocks.} For each of the eight possible blocks, we draw the struts (orange) and hinges (blue) that couple neighboring sides, and mark one of the two polarizations of the soft mode of deformation (top); we suggest a mechanical realization (center), and show its two deformations (bottom). Green and purple frames indicate holographic and non-holographic blocks, respectively.}
\label{fig:hex_blocks}
\end{figure*}

The cubic lattice has three principal directions in three dimensions, and each cubic building block has three pairs of opposite facets. This is clearly related to the square lattice, which has two principal directions in two dimensions, and is made of square blocks, each with two pairs of opposite facets. However, there is another 2D structure that may be compared to the 3D cubic lattice~\cite{pisanty2021putting}: the honeycomb lattice, which also has three principal directions, and its hexagonal building blocks have three pairs of opposite facets. Here too, we assume the same magnitude of deformation on all sides of the block, and use rotations and inversions to reduce the total number of $2^6 = 64$ displacement states to the eight blocks shown in Fig.~\ref{fig:hex_blocks}.

We apply the same analysis, used above for the previous two lattices, in order to classify the possible blocks in the honeycomb lattice according to the deformations along the three principal directions. Although it does have three, like the 3D cubic lattice, the mutual spatial arrangement of these axes is less symmetric in 2D. As a result, two of the cubic block types split to two different hexagonal blocks each, which we denote by Blocks~H$3_a$ and~H$3_b$, and by Blocks~H$5_a$ and~H$5_b$, respectively. Similarly to the cases of the square and cubic lattices, when considering compatible metamaterials in the honeycomb lattice using one of the eight hexagonal building blocks, we may divide them into the same groups: Blocks~H1 and H2 exhibit alternating holographic order, H$3_a$ and H$3_b$ exhibit persistent holographic order (all encircled by the green frame in Fig.~\ref{fig:hex_blocks}), while Blocks~H4, H$5_a$, H$5_b$, and H6 do not exhibit holographic order (encircled by the purple frame in Fig.~\ref{fig:hex_blocks}).

As with the square and cubic lattices, the dashed orange lines denoting struts and the dotted blue lines denoting hinges in Fig.~\ref{fig:hex_blocks} assist in identifying how many distinct orientations each block type has: H1 and H$3_b$ are isotropic, H2 and H$3_a$ each have three possible orientations, and each of Blocks~H4, H$5_a$, H$5_b$, and H6 may be oriented in six possible ways.

\subsection{Symmetries}

In our analysis of each of our three lattices we separated blocks that induce holographic order from those that do not. In all situations, holographic order means that information on the direction of deformation on one facet of the block implies information on the displacement of the diametrically opposite facet. This can be re-phrased as a certain kind of symmetry. For a precise statement, one can turn to hinges and struts: By looking at Figs.~\ref{fig:square_blocks}, \ref{fig:cubic_blocks}, and \ref{fig:hex_blocks}, we see that the distribution of hinges and struts around the boundary is centrally symmetric in exactly the holographic (framed green) cases.

We close this section by pointing out that none of our block types comes in distinct left- and right-handed versions (i.e., they are \emph{achiral}). Indeed, it is easy to check that a reflection (through a line in 2D and through a plane in 3D) of any of our blocks can also be realized by a rotation of the block. In this regard, note that Block~H6, although the motion of its sides may first appear chiral, has a structure in terms of hinges and struts that is clearly achiral.

\section{Mechanical Realization of the Building Blocks}
\label{sec:realization}

We realize all 2D blocks using a unified strategy, which we demonstrate for the square and hexagonal blocks in Figs.~\ref{fig:square_blocks} and~\ref{fig:hex_blocks}. We position at the center of each square or hexagon a smaller, rigid square or hexagon, respectively, that rotates when the soft mode is actuated. We connect inner and outer vertices by radial linkages of equal length. Upon the internal rotation, among the two corners of the internal polygon that face an edge of the block, one moves toward the external edge and one moves away from it. Thus, to make the external edge bend inward, we connect its center to the corner that moves away from that edge, and to make it move outward, we connect it to the corner that moves toward it. This results in each edge of the block having a small, rigid triangle that is based on half of the edge, and which is connected at its opposite corner to one of the corners of the internal polygon. Next to each triangle, in between the same pair of internal and external edges, is a parallelogram. In this way, each square block has four connected floppy parallelograms and each hexagon has six, and upon actuation of the soft mode, all these parallelograms are sheared simultaneously by the same degree but in different directions, depending on the design. 

We note that these deformation modes are not limited to infinitesimal deformations, as is often the case with floppy modes that are obtained only to linear order. Instead, the finite amplitude soft degrees of freedom in these designs serve as mechanisms. Note also that for the designs shown here, with free rotations at contacts between elements, each block's mode of deformation is floppy. That is, there is no mechanical resistance to actuating it, or alternatively no energetic cost for exciting the mode~\cite{chen2014, fruchart2020}. This is in contrast to most realizations of mechanical metamaterials using 3D printing of flexible structures, for which the relevant modes merely have much lower rigidity than other modes of deformation~\cite{coulais2016metacube, meeussen2020supertriangles, Deng_domain_walls_2020, bertoldi_flexible_review_2017, singh_mechanisms_2021}. In the theoretical analysis that follows we will suppress this distinction between floppy (i.e., zero rigidity) and soft (i.e., low rigidity) modes.

Figure~\ref{fig:cubic_blocks} shows how we realize the different 3D cubic blocks. We utilized the fact that each cubic block is made of three orthogonal square blocks. For most of the blocks it is enough to connect two planes and couple the edges in the crossover. This is done best by coupling edges that both move in or both move out. For all cubic blocks besides Block~C3 this is possible, and for Block~C3 we need to have a Block~S3 in all three planes. Therefore, we construct the different cubic blocks by positioning the following square blocks in the planes orthogonal to the $x$, $y$, and $z$ axes: $\mathrm C1=(\mathrm S1, - ,\mathrm S1)$, $\mathrm C2=( - ,\mathrm S2,\mathrm S2)$, $\mathrm C3=(\mathrm S3,\mathrm S3,\mathrm S3)$, $\mathrm C4=(\mathrm S1, - ,\mathrm S4)$, $\mathrm C5=(\mathrm S4,\mathrm S4, - )$, and $\mathrm C6=(\mathrm S2, - ,\mathrm S4)$. Note that the design of the square blocks is slightly modified from those in Fig~\ref{fig:square_blocks} in order to make room for the different planes, see Appendix~\ref{sec:Cubic designl}

\section{Mechanical Compatibility} 
\label{sec:compatibility}

For each of the aforementioned three lattices, we will consider combinatorial metamaterials constructed by positioning each block within the lattice in any one of its possible orientations. We will restrict ourselves to lattices in which all blocks are of the same type, \rs{though} many aspects of the present work may be generalized to lattices \rs{composed of mixed} block types. To test for mechanical compatibility, we will follow loops of successive adjacent blocks within the lattice. Namely, we assume that one block deforms \rs{into} one of its two polarizations, then --- based on the direction of deformation of its facets --- we deduce how its neighboring blocks \rs{must} deform, and continue \rs{this process} until we return to the \rs{original} block. If there is a consistent assignment of polarization for all blocks along this loop, \rs{the deformation is compatible, and} the blocks can all deform in tandem according to their soft mode. If, on the other hand, \rs{we encounter} a contradiction \rs{upon closing the loop, the configuration is said to be} frustrated, \rs{indicating that not all blocks can deform} simultaneously \rs{in} their soft mode.

\rs{This compatibility condition can be formalized using a parity-based approach. In two dimensions, we consider loops surrounding individual vertices. Starting with an arbitrary facet around the vertex, we assign its direction of deformation as $S_0 = 1$ if it deflects clockwise and $S_0 = -1$ if counterclockwise. As we move from one facet to the next, the nature of the connecting corner determines the updated deformation direction: if the corner contains a strut, the direction is preserved ($S_i = S_{i-1}$); if it contains a hinge, the direction flips ($S_i = -S_{i-1}$). After traversing the entire loop (in four steps for the square lattice, and three for the honeycomb), the final deformation is $S = (-1)^h S_0$, where $h$ is the number of hinges in the loop. If $h$ is even, the deformation returns to its original sense ($S = S_0$), and the loop is compatible. If $h$ is odd, the deformation direction is reversed ($S \ne S_0$), indicating that the vertex is frustrated and constitutes a mechanical defect. A similar analysis holds in three dimensions, where the compatibility condition applies to loops encircling lattice edges, with the parity of hinges again determining whether an edge is defect-free or frustrated.}

We will test for compatibility locally. Namely, for each lattice, we will consider all the minimal loops within it. In 2D, a minimal loop is a loop containing the blocks surrounding a vertex in the lattice, and in 3D it contains the blocks surrounding an edge in the lattice. Any larger loop in the lattice may be constructed as a combination of such minimal loops, and therefore, if all the minimal loops are compatible, so is any other loop, too. \rs{More precisely, let us emphasize that in this paper we only consider metamaterials enclosed in simply connected regions; in those cases, in the sense of mod $2$ chains, any loop can indeed be written as a sum of elementary loops. Furthermore, if we associate $0$ to compatible loops and $1$ to incompatible ones, we obtain a mod $2$ valued mapping that is additive on all loops.}

\begin{figure}[t]
\centering
\includegraphics[width=0.8\columnwidth]{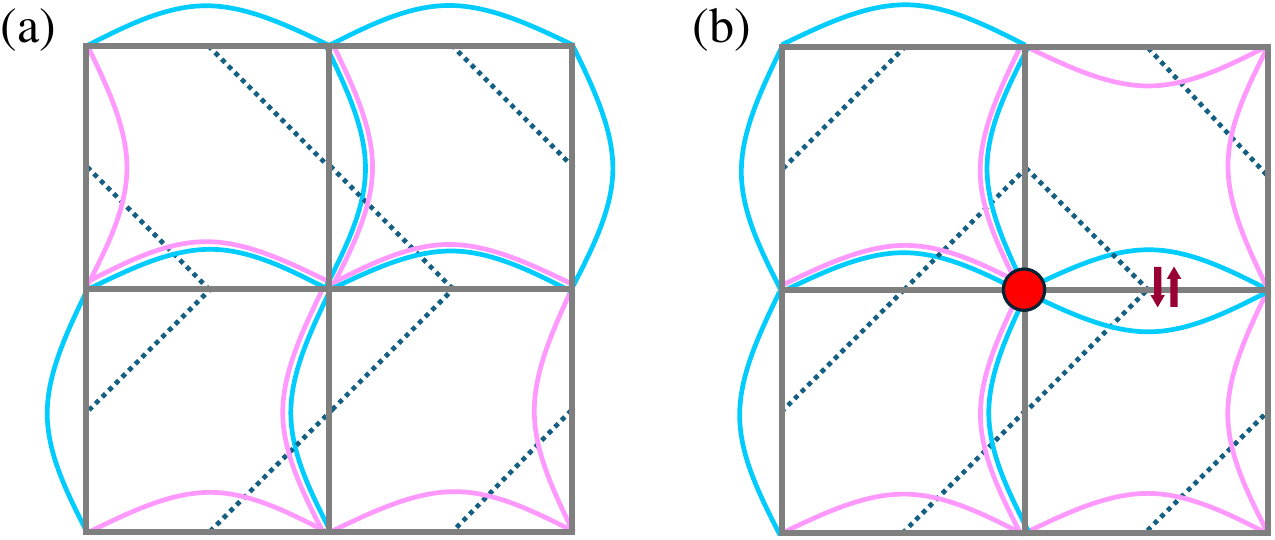}
\caption{\textbf{Mechanical compatibility.} Four Block~S3 square units can be mutually oriented, (a) in a compatible manner such that they can all simultaneously deform in their soft mode, or (b) in a frustrated manner in which there must be at least one block that cannot deform in its soft mode. The red dot indicates the defected vertex at the core of the frustrated loop.}
\label{fig:compatibility}
\end{figure}

Figure~\ref{fig:compatibility} demonstrates mechanical compatibility for a minimal loop in the square lattice, which includes four building blocks around a vertex in the lattice. Here, we use Block~S3 to show that certain orientations of the blocks give a compatible structure, while other orientations lead to frustration. In compatible metamaterials, all loops are compatible. If a minimal loop in 2D is frustrated, we can identify the vertex that this loop encircles as a mechanical defect in the lattice. In 3D, minimal loops encircle edges in the lattice, and these edges will be the basic objects that are either compatible or frustrated. In this paper we will focus on compatible metamaterials, while in an accompanying paper we discuss frustrated situations with defects within the lattice~\cite{defects_paper}.

\section{Multiplicity of Compatible Structures} 
\label{sec:multiplicity}

In this section we derive results regarding the number, also referred to as the \emph{multiplicity}, of compatible metamaterials that may be constructed from the various block types. Table~\ref{results_table} lists the block types introduced in Sec.~\ref{sec:blocks} above and summarizes the findings derived in the present section. It is meant to enable readers to skip, or alternatively focus on, the technical parts of this section where the various results are established. `Holography' indicates whether the block induces alternating or persistent holographic order, or does not induce holographic order at all. `Orientations' are the number of distinct orientations that each block has. `Multiplicity' indicates how the number of compatible metamaterials that may be constructed from each block type depends on system size. From each of the isotropic blocks, only one metamaterial may be constructed, and depending on the block type, that metamaterial is either compatible or not, thus trivially yielding either 1 or 0 compatible metamaterials. For all the anisotropic blocks, we find that the multiplicity of compatible metamaterials grows exponentially either with the number of blocks in the lattice or with the number of blocks along the boundary of the system. In analogy to statistical mechanics, where entropy is related to the logarithm of the number of states, we will refer to the former case as `extensive' and to the latter as `sub-extensive'. 

\begin{table}[t]
\begin{center}
\center
\begin{tabular}{| c | c | c | c | c |}
\hline
Block & Holography & Orientations & Multiplicity \\
\hline
\hline
S1     & Alternating & 1  & 1             \\
S2     & Alternating & 1  & 1             \\
\hline
S3     & Persistent & 2  & sub-extensive \\
\hline
S4     & No  & 4  & extensive     \\
\hline
\hline
C1     & Alternating & 1  & 1             \\
C2     & Alternating & 3  & sub-extensive \\
\hline
C3     & Persistent & 4  & sub-extensive \\
\hline
C4     & No  & 6  & extensive \\
C5     & No  & 12 & extensive \\
C6     & No  & 6  & extensive \\
\hline
\hline
H1     & Alternating & 1  & 0 \\
H2     & Alternating & 3  & sub-extensive \\
\hline
H3$_a$ & Persistent & 3  & sub-extensive \\
H3$_b$ & Persistent & 1  & 1 \\
\hline
H4     & No  & 6  & extensive \\
H5$_a$ & No  & 6  & extensive \\
H5$_b$ & No  & 6  & extensive \\
H6     & No  & 6  & extensive \\
\hline
\end{tabular}
\end{center}
\caption{Summary of the symmetries of the various building blocks and the resulting behavior of metamaterials built from them.} 
\label{results_table}
\end{table}

For the 2D square lattice, we consider square regions of $L \times L$ square blocks. For the 2D honeycomb lattice, we consider hexagonal regions with linear size $L$ that include $3L^2-3L+1$ hexagonal blocks. Finally, for the 3D cubic lattice, we consider cubic regions of $L \times L \times L$ cubic blocks. For each type of block, we will ask what the number $\Omega(L)$ of different compatible metamaterials is, that may be constructed using the given block. Specifically, we will be interested in how this number scales with $L$, which is the linear size of the system. We refer to cases in which this number grows exponentially with the number of blocks in the lattice as extensive. In such cases, $\Omega \sim \exp(L^3)$ in 3D, and $\Omega \sim \exp(L^2)$ in 2D. In other cases, which we refer to as sub-extensive, we find that $\Omega \sim \exp(L^2)$ in 3D and $\Omega \sim \exp(L)$ in 2D, in particular $\Omega(L)$ grows more slowly with system size, and grows exponentially only with the size of the system's boundary. Previous work has shown that for the cubic lattice of Block~C2~\cite{coulais2016metacube} as well as for the honeycomb lattice of Block~H2~\cite{pisanty2021putting}, the scaling is sub-extensive. As explained below, for blocks that induce holographic order, the multiplicity of compatible metamaterials is generally sub-extensive. For those that do not induce holographic order, even though we are not aware of a general argument, we do find that for all the block types that we consider here, the scaling is extensive.

\subsection{Isotropic Blocks}

\begin{figure}[t]
\centering
\includegraphics[width=.75\columnwidth]{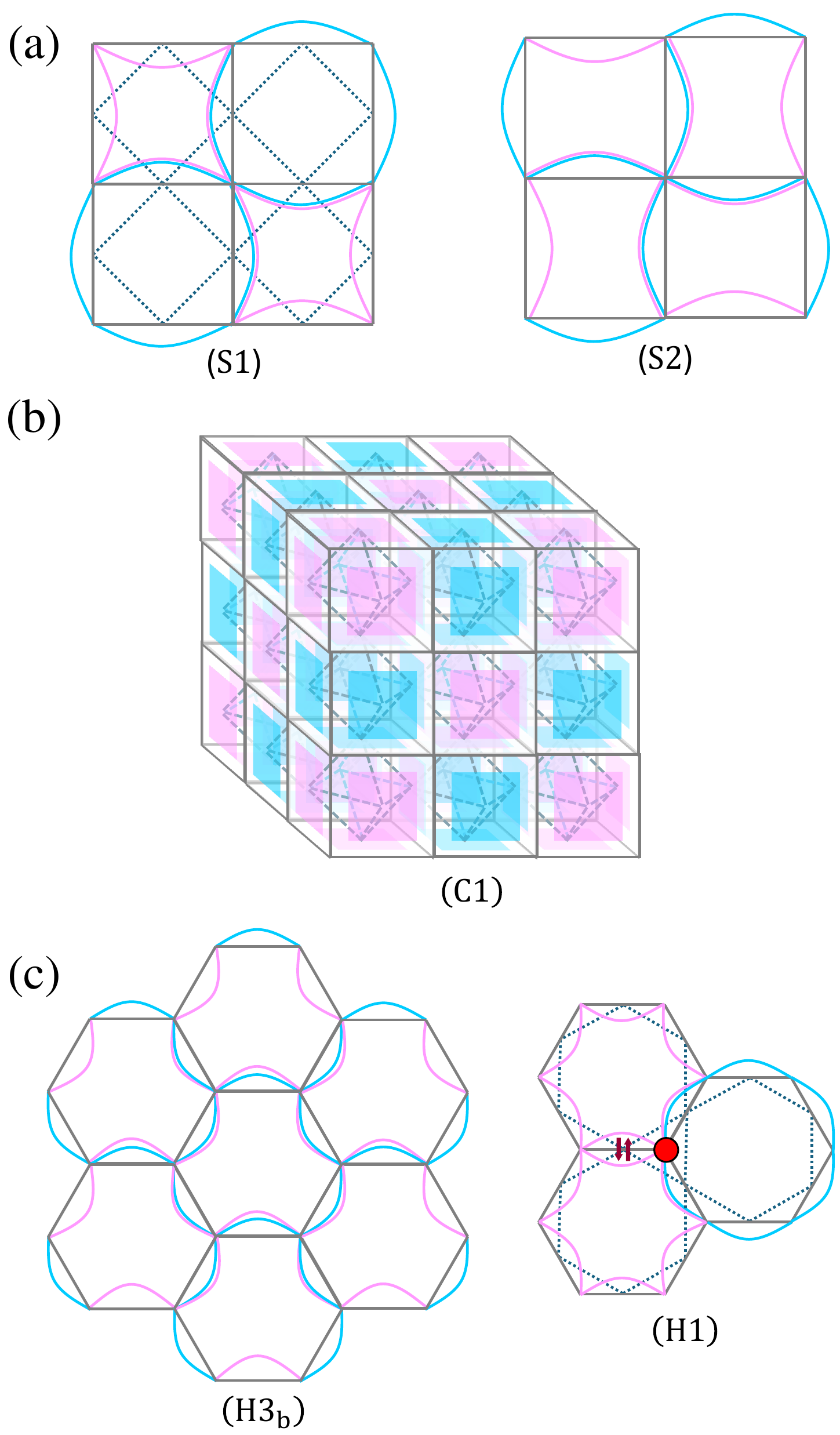}
\caption{\textbf{Isotropic blocks.} Unique compatible structures made of Block~S1 and~S2 in the square lattice (a) and of Block~C1 in the cubic lattice (b). Tiling the honeycomb lattice with Block~H$3_b$ yields a compatible metamaterial, while tiling it with Block~H1 leads to mechanical frustration (c).}
\label{fig:isotropic}
\end{figure}

The structure of Blocks~S1, S2, C1, H1 and H$3_b$ is isotropic. Thus, from each one of them, there is only one way to construct a metamaterial. The square and the cubic lattices are bipartite, and metamaterials made of the isotropic blocks in these lattices allow all blocks to deform in their preferred form, as shown in Fig.~\ref{fig:isotropic}a,b, respectively, by having alternating polarizations of neighboring blocks. Therefore, these metamaterials are mechanically compatible, and for these blocks, the multiplicity is the trivial $\Omega_{\textrm{S}1} = \Omega_{\textrm{S}2} = \Omega_{\textrm{C}1} = 1$. In the honeycomb lattice, on the other hand, as depicted in Fig.~\ref{fig:isotropic}c, a metamaterial built of Block~H$3_b$ is compatible, thus $\Omega_{\textrm{H}3_b} = 1$, while one made of Block~H1 is not, that is $\Omega_{\textrm{H}1} = 0$.

\subsection{Anisotropic Holographic Blocks}

\begin{figure}[t]
\centering
\includegraphics[width=\columnwidth]{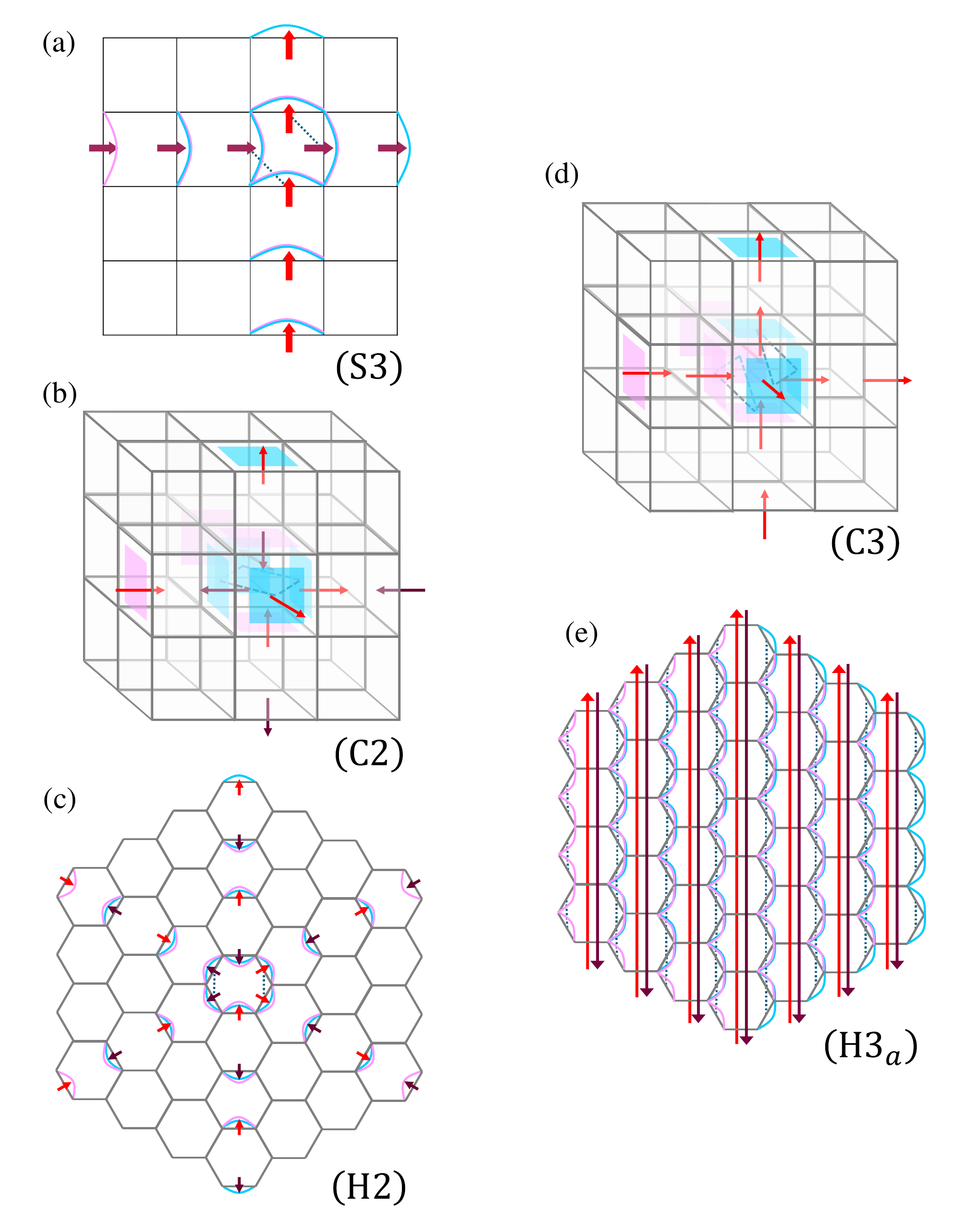}
\caption{\textbf{Holographic blocks.} (a) The deformation texture on two adjacent sides of a metamaterial made of Block~S3 determines the orientations of all internal blocks. The orientation of Blocks C2~(b) and H2~(c) in the bulk are governed by the texture on the boundary. (d) The deformation texture on three adjacent faces of a metamaterial made of Block~C3 determines the orientations of all internal blocks. (e) Sub-extensive bound for Block~H$3_a$.}
\label{fig:holographic}
\end{figure}

Block~S3 can be positioned in two possible orientations, thus for a square lattice of $L \times L$ blocks, there is a total of $2^{L^2}$ metamaterials that can be constructed from it. However, only some of these allow all blocks to simultaneously deform in their preferred mode, and are thus mechanically compatible. Due to the holographic order that Block~S3 induces, any displacement texture on two adjacent sides of the metamaterial uniquely determines the displacement field inside the metamaterial, such that all blocks within the metamaterial deform according to their preferred soft mode, as shown in Fig.~\ref{fig:holographic}a. Moreover, all compatible metamaterials may be constructed in such a manner starting from a displacement texture on two adjacent sides of the metamaterial. Thus, the number of compatible metamaterials made of Block~S3 is equal to $\Omega_{\textrm{S}3} = 2^{2L-1}$, where we have divided by two because each compatible metamaterial has two polarizations of deformation textures that differ by an overall inversion of all displacements.

As with the square lattice made of Block~S3, also with the non-trivial holographic Blocks~C2 and C3 in the cubic lattice and Blocks~H2 and H$3_a$ in the honeycomb lattice, the deformation texture on the boundary determines the deformations everywhere within the bulk of the system. This trivially leads to a sub-extensive upper bound on the multiplicity of compatible metamaterials, which scales with the possible freedom on the boundary of the system. However, generally not all deformation textures on the boundary yield for all blocks within the system deformations that are consistent with the soft mode of the blocks at hand. Therefore, the actual multiplicity of compatible metamaterials may be lower than this bound. 

As shown in Fig.~\ref{fig:holographic}b,c, compatible metamaterials are obtained from Blocks~C2 and H2 only for boundary textures that, for every block within the system, lead to motion out of the block or in to it along two axes and to motion in the opposite direction for the third axis. In other words, the boundary conditions should not force any block to behave in the manner of C1 or H1. This condition implies a lower bound on the number of compatible metamaterials, as well as a tighter upper bound, which both scale sub-extensively with system size. Therefore the multiplicity of compatible metamaterials made of these blocks scales exponentially with the system's boundary, i.e., $\Omega_{\textrm{C}2} \propto \exp (L^2)$ and $\Omega_{\textrm{H}2} \propto \exp (L)$. For the complete derivation of these bounds and for the precise values of the coefficients, see~\cite{coulais2016metacube, pisanty2021putting}. 

We will now consider the remaining non-trivial holographic cases -- the cubic lattice made of Block~C3 and the hexagonal lattice made of Block~H$3_a$.

Block~C3 behaves very much like Block~S3 in the square lattice in that the motion of any of its faces into the block implies motion of the opposite face out of the block. Therefore, any deformation texture on three adjacent faces of the lattice will lead to deformations for all building blocks within the lattice that are consistent with this soft mode, see Fig.~\ref{fig:holographic}d. Thus the number of compatible metamaterials made of Block~C3 is $\Omega_{\textrm{C}3} = 2^{3L^2-1}$, where as with Block~S3, we divided the number of boundary textures by two since each compatible metamaterial has two possible polarizations of its global soft mode of deformation.

As with the other holographic blocks, an upper bound on the number of compatible metamaterials made of Block~H$3_a$ is obtained from the freedom on its boundary. For a hexagonal system with $L$ blocks along each of its sides, in each of the three principal directions there are $2L-1$ lines that run through the lattice. Since along each of these lines the deformation can take two possible directions, an upper bound on the number of compatible metamaterials is $\Omega_{\textrm{H}3_a} \le 2^{6L-4}$, where we have divided the number of possible deformation states by two since each compatible metamaterial has two polarizations with opposite deformations. To obtain a lower bound, we note that if we fix the deformations in the same direction along two axes as shown in Fig.~\ref{fig:holographic}e, we can independently set the direction of deformation along all the $2L-1$ lines running through the system along the third axis. This leads to $\Omega_{{H}3_a} \ge 2^{2L-2}$. 

We conclude that for all anisotropic holographic building blocks, the number of compatible structures grows sub-extensively with system size.

\subsection{Non-Holographic Blocks}

Block~S4 has four possible orientations, thus for an $L \times L$ lattice, there is a total of $4^{L^2}$ possible metamaterials that may be constructed from it. This upper bound $\Omega_{\textrm{S}4} < 4^{L^2}$ on the number of compatible metamaterials scales extensively with system size, i.e., exponentially with the number of blocks in the lattice. We will now establish an extensive lower bound on the number of compatible metamaterials made of Block~S4. Since both these bounds scale similarly with system size, we will deduce that the multiplicity of compatible metamaterials made of Block~S4 is also extensive.

\begin{figure}[t]
\centering
\includegraphics[width=0.8\columnwidth]{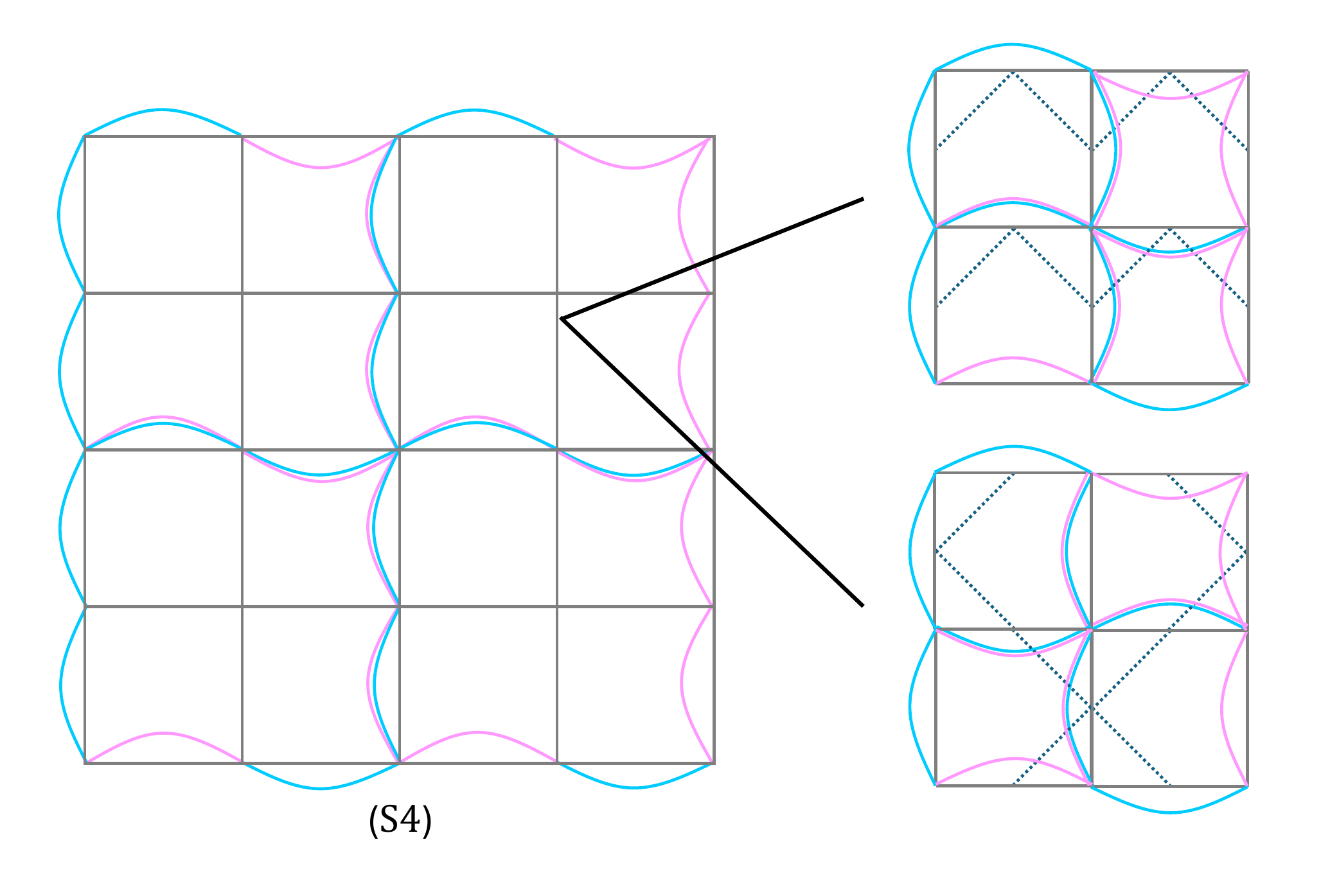}
\caption{\textbf{Non-holographic square block.} Two possible arrangements (right) of Blocks~S4 give super-blocks that have the same deformation texture on their boundary, and which can tile the plane (left).}
\label{fig:supercellsSqr}
\end{figure}

To obtain a lower bound, we consider the two \emph{super-blocks}, each made of four blocks and shown in Fig.~\ref{fig:supercellsSqr}. Both of the super-blocks have the same behavior on their boundary, and this behavior allows to tile the plane with these super-blocks in a mechanically compatible manner. However, the orientations of the blocks forming these two super-blocks are different. Thus, we can construct a metamaterial via a regular tiling by these super-blocks, but with an arbitrary choice of which of the two super-blocks will be used in each repetition in the lattice. In an $L \times L$ lattice, there are $L^2/4$ such super-blocks, which can be used to form $2^{L^2/4}$ different compatible metamaterials. There are, of course, additional compatible metamaterials made of Block~S4, and this construction establishes only a lower bound, $\Omega_{\textrm{S}4} > 2^{L^2/4}$, but it is, as stated above, extensive. 

\begin{figure}[t]
\centering
\includegraphics[width=.8\columnwidth]{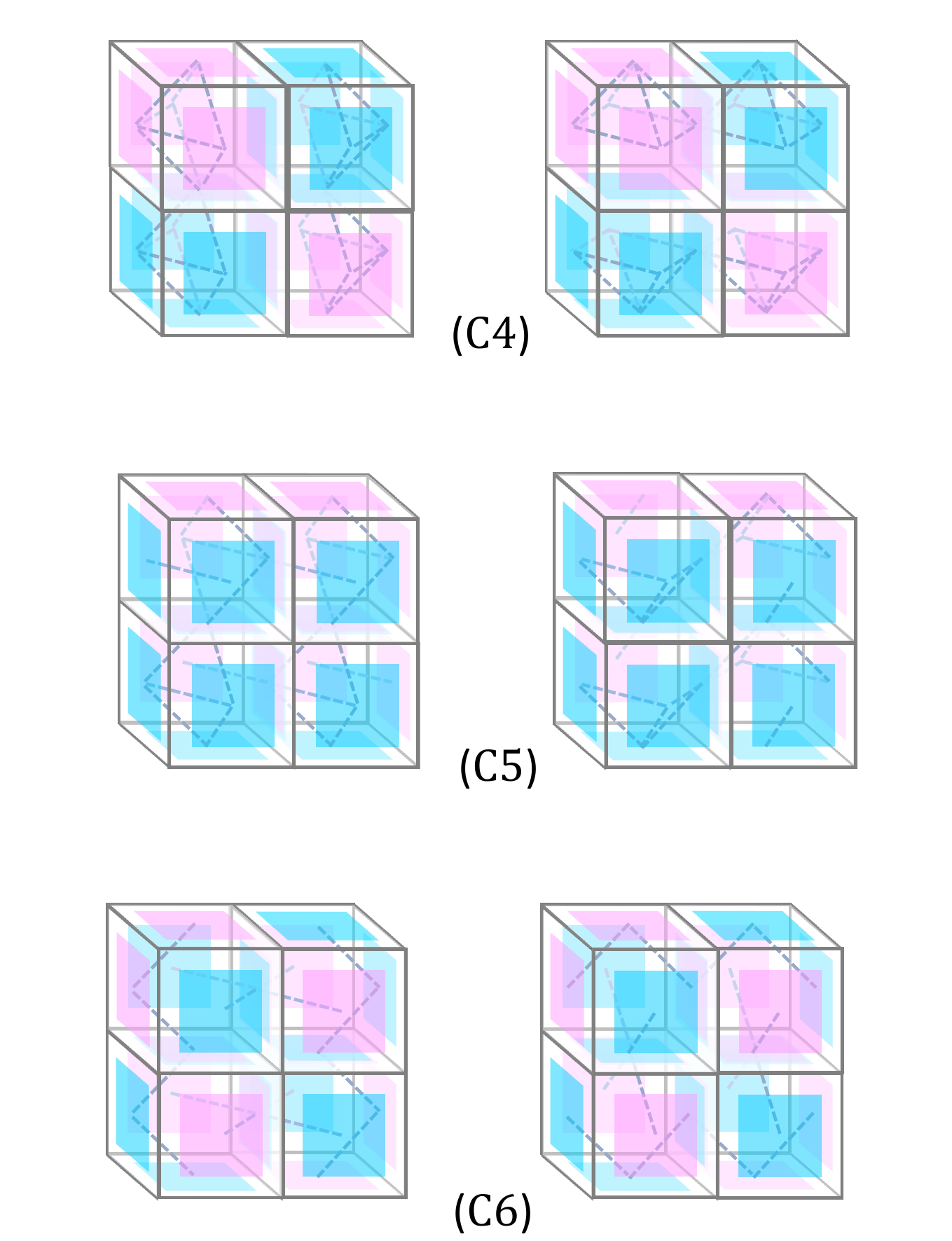}
\caption{\textbf{Non-holographic cubic blocks.} Periodic super-blocks that have two internal configurations of the building blocks with the same outer texture, for Blocks~C4, C5 and C6.}
\label{fig:supercells3d}
\end{figure}

\begin{figure}[t]
\centering
\includegraphics[width=\columnwidth]{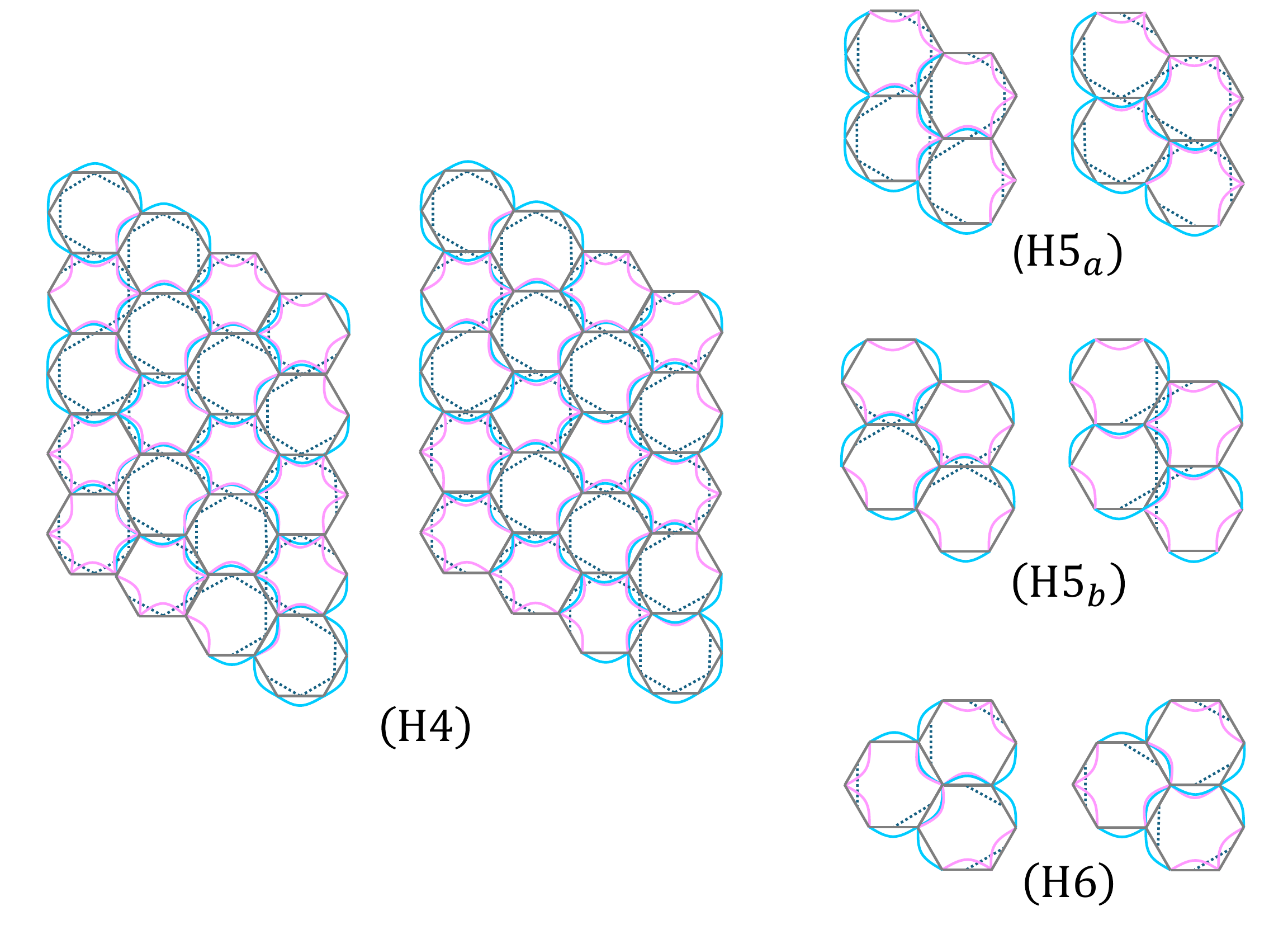}
\caption{\textbf{Non-holographic hexagonal blocks.} Periodic super-blocks that have two internal configurations of the building blocks with the same outer texture for Blocks~H4, H$5_a$, H$5_b$ and~H6.}
\label{fig:supercells2d}
\end{figure}

\rs{For Block~S4 our approach to design the deformation texture on the boundary of the metamaterial, as presented in Sec.~\ref{sec:texture} below, enables us to derive an exact expression for $\Omega_{\textrm{S}4}$. However, the bounds presented above already constitute} our first demonstration of how the absence of holographic order allows extensive multiplicity of compatible metamaterials. Similarly to the approach employed above, in Figs.~\ref{fig:supercells3d} and~\ref{fig:supercells2d}, we respectively present super-blocks that tile space for the non-holographic Blocks~C4, C5, and C6 in the cubic lattice, as well as \rs{plane-tiling super-blocks made of} Blocks~H4, H$5_a$, H$5_b$, and H6 in the honeycomb lattice.

Blocks~C4 and C6 can be oriented in six possible directions, and Block~C5 has 12 possible orientations. Thus, the total number of metamaterials of size $L \times L \times L$ gives the upper bounds $\Omega_{\textrm{C}4} < 6^{L^3}$, $\Omega_{\textrm{C}5}< 12^{L^3}$, and $\Omega_{\textrm{C}6} < 6^{L^3} $ for the number of compatible metamaterials made of each one of these three blocks. In all these three cases, the super-blocks shown in Fig.~\ref{fig:supercells3d} are made of four blocks. Therefore one may place $L^3/4$ such super-blocks in the lattice, and since each one of them can be of one of the two types presented here, we obtain the lower bounds $\Omega_{\textrm{C}4} , \Omega_{\textrm{C}5} , \Omega_{\textrm{C}6} > 2^{L^3/4}$. As both the upper and lower bounds scale extensively with system size, so does the actual multiplicity.

Hexagonal Blocks~H4, H$5_a$, H$5_b$ and H6 all have six possible orientations, leading to the upper bound $\Omega_{\textrm{H}4} , \Omega_{\textrm{H}5_a} , \Omega_{\textrm{H}5_b} , \Omega_{\textrm{H}6} < 6^N$, where $N$ is the number of blocks in the lattice, which for a hexagon of side $L$ is given by $N=3L^2-3L+1$. The super-blocks made of Blocks~H4, H$5_a$, H$5_b$ and H6, shown in Fig.~\ref{fig:supercells2d}, are made of 20, 4, 4, and 3 blocks, respectively. These super-blocks were found using a computer program we wrote that finds compatible structures made up of the various hexagonal blocks~\cite{meta_hex_program}. Therefore, lower bounds for the number of compatible metamaterials made of these blocks are given by $\Omega_{\textrm{H}4} > 2^{N/20}$, $\Omega_{\textrm{H}5_a} > 2^{N/4}$, $\Omega_{\textrm{H}5_b} > 2^{N/4}$, and $\Omega_{\textrm{H}6} > 2^{N/3}$. As with the non-holographic square and cubic blocks, the extensive scaling with system size of both the lower and the upper bound allows us to conclude that metamaterials made of the non-holographic hexagonal blocks have extensive multiplicity, too.

\section{Texture Design}
\label{sec:texture}

One of the exciting features of combinatorial mechanical metamaterials is that they enable the design of deformation textures~\cite{coulais2016metacube, bertoldi_flexible_review_2017}. Namely, the theoretical understanding of the combinatorial matching rules between the building blocks comprising the metamaterial enables one to design their orientations in such a way that the global soft mode of deformation of the system has a desired pattern on the boundary of the metamaterial. Here we can think of the deformation texture along a face of a 3D cubic metamaterial as a 2D matrix of binary pixels, where each pixel prescribes whether the face of the block at that position moves into or out of the metamaterial. We will represent the two possibilities with two different colors. We will ask whether any such pixelized image may be realized, and furthermore whether arbitrary images may be simultaneously realized on all six faces of a cubic metamaterial. 

For a 2D metamaterial, we will similarly consider each 1D side of the metamaterial, and define the deformation texture on that side as a 1D array of pixels, which specifies whether the part of each block along that side moves into or out of the metamaterial. In the honeycomb lattice, each side of the hexagonal domain contains facets in two of the principal directions of the lattice, and even designing the texture on a single side of the lattice is not obviously possible~\cite{pisanty2021putting}. For that reason we will limit the following discussion in 2D to the square lattice.

We note that inverting every pixel along the boundary, both in 2D and 3D cases, describes the same boundary texture as before, just in the opposite polarization of the desired global soft mode that realizes the texture. In other words, a \emph{texture} is equivalent to a pair of two \emph{colorings}, related by exchanging the colors.

Following the arguments presented above on the multiplicity of compatible metamaterials, with Block~S3 we can design the texture on any two adjacent sides of the square lattice, and with Block~C3 we can design the texture on any three adjacent faces of the cubic lattice. Clearly, due to the holographic nature of these blocks, the texture on the opposite facets is fully determined and may not be independently assigned. For the other holographic blocks, and specifically for those exhibiting alternating holographic order, things are more complicated. 

For example, using Block~C2, an arbitrary texture can be assigned to a single face of the metamaterial~\cite{coulais2016metacube}. However, for this block type, the deformation textures on the other two pairs of faces cannot be independently assigned, either. This may be explained using the following counting argument. Realizing a surface texture requires at least one choice of the orientations of all blocks everywhere within the metamaterial. So, the number of surface textures that may be realized cannot be larger than the multiplicity of compatible metamaterials. For a metamaterial made of $L \times L \times L$ blocks, the number of possible colorings on each of its faces is $2^{L^2}$, and therefore for its three pairs of opposite faces this number is $2^{3L^2}$. Thus by the remark above, we have $2^{3L^2-1}$ texture candidates. However, for large $L$, the multiplicity of compatible metamaterials is smaller than $4L \left( \frac{3}{4} \right)^L 2^{2L^2}$, and exact enumeration up to $L=14$ indicates that the actual values are closer to the lower bound of $3 \cdot 2^{L^2+L}$~\cite{coulais2016metacube}. The established upper bound therefore scales as the number of possible textures on two faces, while numerical results suggest a rate of growth that scales as the number of possible textures on one face. In any case, it is clear that three faces cannot be simultaneously designed.

We will now analyze metamaterials made of square Block~S4 and those made of cubic Block~C5 to demonstrate how, both in 2D and in 3D, non-holographic blocks allow for the design of a soft mode of the metamaterial such that it will have any desired deformation texture on its entire boundary, except for a single pixel, which is not free due to a parity constraint. 

\subsection{Square Lattice}
\label{ssec:square}

Block~S4 does not impose holographic order. Therefore, we could attempt to build from it a compatible metamaterial with a designed deformation texture on its entire boundary. However, there is a parity constraint on the numbers of in and out pixels on the boundary of the metamaterial. Let us consider a lattice of $L_x$ by $L_y$ blocks, whose boundary consists of $2(L_x+L_y)$ pixels. We also consider a compatible metamaterial in the lattice, and `freeze' it in one of its polarizations. If we denote by $B_i$ the number of in pixels along the boundary, then the number of out pixels will be $B_o=2(L_x+L_y)-B_i$. If $N_i$ blocks deform with 3 sides in and 1 side out, then $N_o=L_x L_y-N_i$ blocks deform with 3 sides out and 1 side in. Now, there are $I = L_x(L_y-1)+L_y(L_x-1) = 2 L_x L_y - L_x - L_y$ internal edges within the bulk of the lattice. Each such internal edge contributes one side into one of the blocks sharing that edge, as well as one side out of the other block, so this is the number of internal ins and also of internal outs. The number of in pixels along the boundary is thus $B_i = 3 N_i + N_o  - I$, where we summed the 3 or 1 ins that the blocks in the two different states contribute, and subtracted the ins at the internal edges. After substitution, this may be written as $B_i = 2 N_i - (L_x-1)(L_y-1) + 1$. In particular, if both $L_x$ and $L_y$ are even then $B_i$ is even, otherwise $B_i$ is odd. Since $B_i+B_o$ is even, $B_o$ has the same parity as $B_i$. A boundary texture with in and out pixels whose parity differs from this will clearly not be realizable. On the other hand, we will now show that any boundary texture that satisfies this parity constraint may be realized. This implies that a boundary texture with the wrong parity may be realized up to a single pixel, which would have to be flipped in order to obtain the correct parity.

\begin{figure}[t]
\centering
\includegraphics[width=0.7\columnwidth]{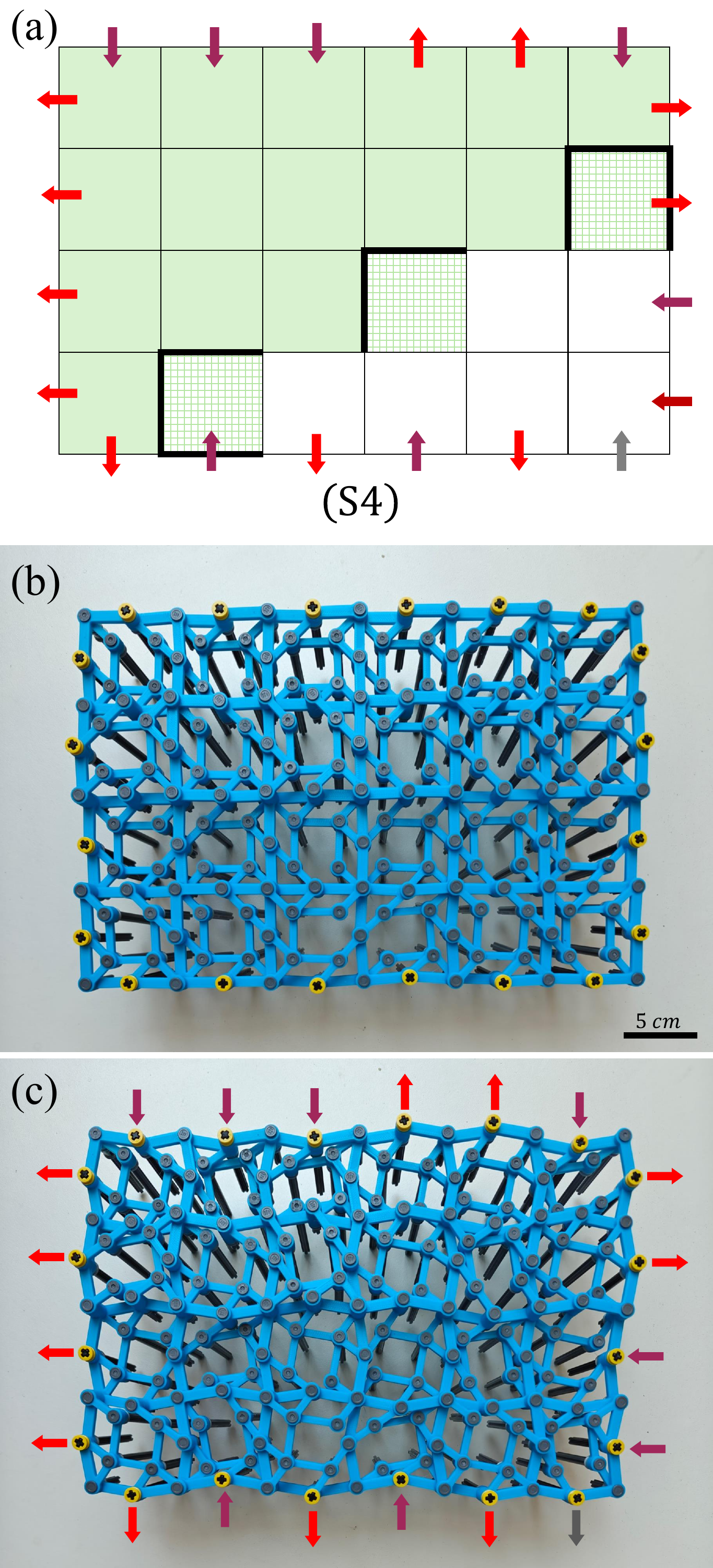}
\caption{\textbf{Texture design in the square lattice.} Red arrows denote the desired texture, while the gray arrow indicates deformation that is determined due to parity. a) Up to three of the sides of each Block~S4 within the lattice, when it is assigned, are set (thick black lines in the three hashed examples), and therefore the block always has at least one possible orientation, except for the last block at the corner, where the displacement on one of its sides is set by parity. \rs{b-c) Experimental demonstration: b) Undeformed lattice. c) Floppy mode with the desired texture. Note that the blocks have a slightly different mechanical design than that given in Fig.~\ref{fig:square_blocks}, and the gray arrow represents the pixel that is determined by parity.}}
\label{fig:Texture_square}
\end{figure}

We scan the lattice systematically and assign block orientations. Since we set the deformation texture on all four sides of the metamaterial, for each block which is reached, the deformations on either two or three sides are set, except for the last block, for which all four sides are set, see Fig.~\ref{fig:Texture_square}\rs{a}. Even if the deformations on three sides are set, there is a way to orient Block~S4 so that this deformation is consistent with its floppy mode. As to the last block, when the aforementioned parity constraint is met, it guarantees that a suitable orientation (uniquely) exists. However, if the parity of desired in and out pixels on the boundary of the metamaterial is inconsistent with the parity constraint, the last block reached in the scanning process will require either that all its four sides move in, or that all move out, or that there are two sides of each type. Any of these is incompatible with the floppy mode of Block~S4, and one last pixel will not move in the desired direction. \rs{In Fig.~\ref{fig:Texture_square}b,c, we supplement this theoretical procedure with an experimental demonstration of a metamaterial that deforms to a predetermined boundary texture.}

\rs{We can now use our theoretical analysis to get an exact expression for the multiplicity of compatible metamaterials that may be constructed from Block~S4: An $L_x \times L_y$ lattice has $2(L_x+L_y)$ boundary pixels. One of these pixels will be set by parity, and all others may independently take one of two values. Thus, after further dividing by $2$ to account for the two opposite polarizations, we see that there are $2^{2(L_x+L_y)-2}$ realizable boundary textures. For each of these possible boundary textures, as we scan the lattice to decide on the orientations of the blocks, for the $L_x+L_y-1$ blocks that are on the bottom and right boundary of the lattice, three (or four, for the block in the bottom-right corner) sides are set, and there is only one possible orientation for the block, while for the remaining $(L_x-1)(L_y-1)$ blocks, only two sides are set and we have two possible orientations per block. Therefore, each boundary texture may be realized by $2^{(L_x-1)(L_y-1)}$ distinct metamaterials. So all together, the multiplicity of compatible metamaterials constructed from Block~C4 is $\Omega_{S4}= 2^{2(L_x+L_y)-2}\cdot 2^{(L_x-1)(L_y-1)} = 2^{L_xL_y +L_x +L_y-1}$. For $L_x=L_y=L$ this reduces to $\Omega_{S4}=2^{L^2+2L-1}$, which is indeed between the bounds given in Sec.~\ref{sec:multiplicity} of $2^{L^2/4}<\Omega_{S4}<2^{2L^2}$.}

\subsection{Cubic Lattice}

Using Block~C5 one can design the texture on all six faces of the lattice, except for one pixel, which is set by parity. \rs{The details of the argument are deferred to a pair of Appendices. Considering a lattice of $L_x \times L_y \times L_z$ blocks, in Appendix~\ref{sec:C5_p} we show that the number of in (and equivalently, out) pixels is even unless exactly one of the three lengths $L_x$, $L_y$ or $L_z$ of the lattice is even.} Assuming this parity constraint holds, we can design any texture on the entire boundary. Block~C5 has two pairs of opposite faces that deform in the same direction, i.e., one in and one out, and one pair that both deform in or both out (see Fig.~\ref{fig:cubic_blocks} above). Therefore, for any desired deformation pattern of four adjacent faces (meaning that the two non-specified faces are adjacent), there is an orientation of Block~C5 that can satisfy it. However, a problem occurs when trying to satisfy two pairs of opposite faces that both deform in to (or out of) the block. We can avoid this situation by properly orienting the blocks that are close to the boundary of the lattice, as described in the protocol detailed in Appendix~\ref{sec:c5 protocall}.

\begin{figure}[t]
\centering
\includegraphics[width=0.95\columnwidth]{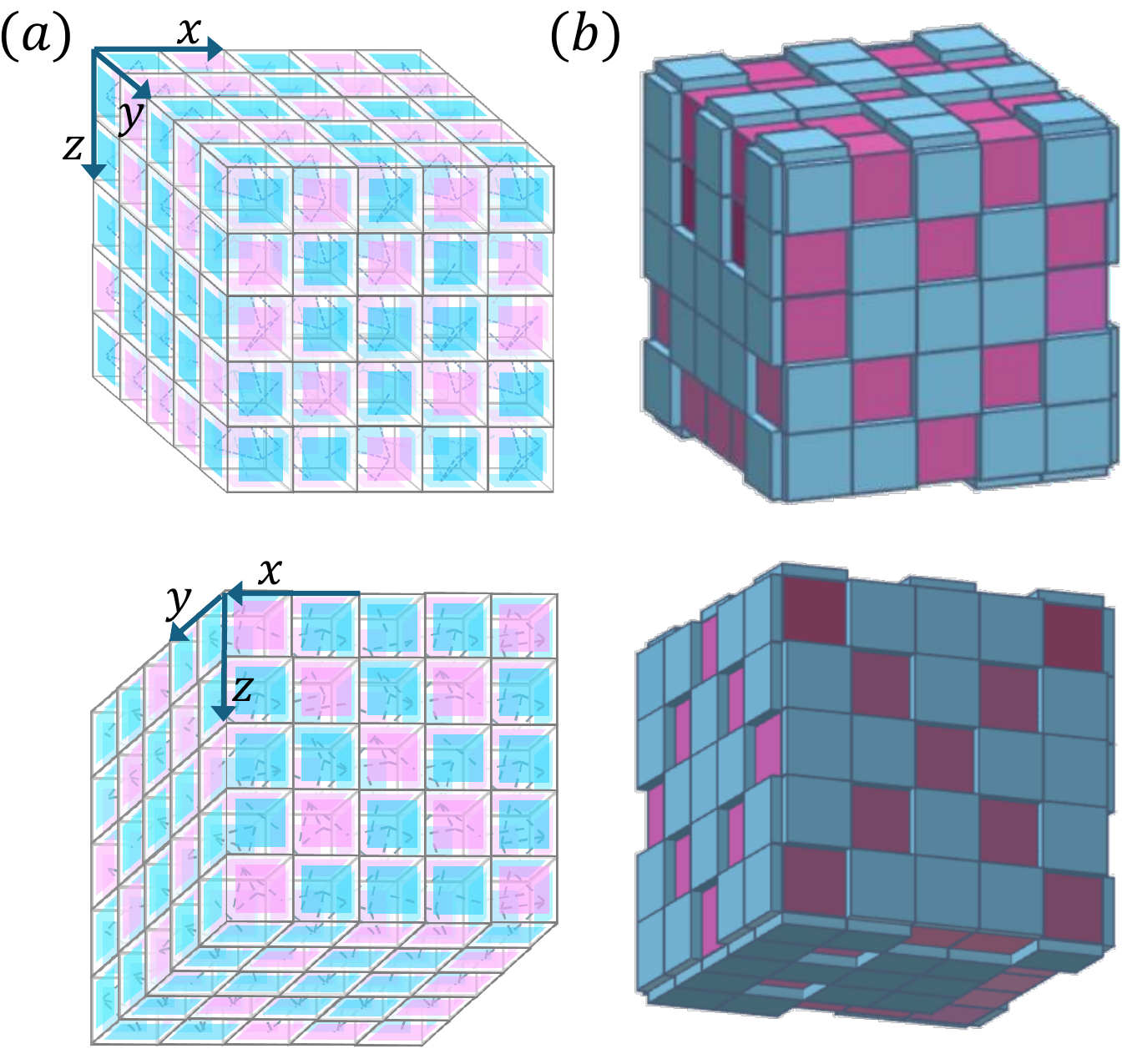}
\caption{\textbf{Texture design in the cubic lattice.} Orientations of all C5 blocks within the lattice (a), yielding the floppy mode specified on all six faces of the cube (b).}
\label{fig:C5_box}
\end{figure}

We have implemented this procedure in a computer program that receives any desired texture on all six faces of the lattice, and outputs the orientations of all the blocks in the metamaterial so that the floppy mode of the system will correspond to the desired boundary texture~\cite{texture_design_code}. Figure~\ref{fig:C5_box} shows a desired deformation texture on all six faces of the cube, and the orientations of all blocks within the bulk that yield this boundary texture, as obtained from our algorithm.

\section{Discussion}
\label{sec:discussion}

We systematically studied combinatorial mechanical metamaterials, constructed using all possible building blocks that possess a simple soft mode of deformation. We showed how for blocks that induce holographic order, the multiplicity of compatible metamaterials scales sub-extensively with system size, while for those that do not induce holography, the scaling is extensive. For non-holographic blocks we demonstrated how we could design the deformation texture on the entire boundary of the system. It would be interesting to see to what extent the deformation texture within the bulk of the system may be designed. The concept of designed (meta)materials with anisotropic blocks could be employed in other physical contexts. Specifically, it would be interesting to see how the effect of holography on the design freedom can enrich the physical behavior of systems ranging from origami~\cite{dieleman_jigsaw_2020}, through artificial magnetic~\cite{wang2006, nisoli2013} or colloidal~\cite{libal2006, ortiz-ambriz2016, oguz2020} systems, to other applications in chemistry and materials science.

We limited ourselves to metamaterials made of building blocks with a single soft or floppy mode of deformation per block. In this case, when the blocks are positioned in a compatible manner, the metamaterial has one collective floppy mode. Using blocks with more than one floppy mode can lead to metamaterials with multiple floppy modes, and more interestingly, the number of modes in the collective system can depend on the orientations of the blocks within the metamaterial~\cite{bossart_oligomodal_2021, van_mastrigt_machine_2022, van_mastrigt_emergent_2023}. Our approach of identifying blocks with geometrically simple modes of deformation, as well as distinguishing between holographic and non-holographic blocks and modes, could be extended to such multimodal blocks.

Our analysis benefited from the fact that we considered blocks with simple modes of deformation which could hence be described in a binary way of whether each facet moves into the block or out of it. It would be interesting to compare this ideal situation to blocks with more complex displacements, either of different amplitudes on different facets or with displacements that are not normal to the facets.

\begin{acknowledgments}

We thank Corentin Coulais, Martin van Hecke, Ben Pisanty, Priyanka, Camilla Sammartino, Dor Shohat, Tomer Sigalov, Ivan Smalyukh, Shai Sonnenreich and Nadiv Swidler for helpful discussions. This research was supported in part by the Israel Science Foundation Grant Nos. 1899/20. Support was also provided by the Japan Society for the Promotion of Science (JSPS) Grant-in-Aid for Scientific Research C, no.\ 23K03108. C.S.K. was supported by the Clore Scholars Programme.

\end{acknowledgments}

\emph{Author Contributions:} C.S.K., T.K. and Y.S. conceived of and planned the research, performed the theoretical research, and wrote the paper; C.S.K. designed and constructed the experimental models; O.P. wrote and used the computer program for studying the honeycomb lattice with different block types and implemented the protocol for texture design with Block~C5; N.S. contributed to classifying the 3D blocks and set the grounds for their experimental realization.

\appendix

\section{Physical construction of the 3D cubic blocks}
\label{sec:Cubic designl}

\begin{figure*}[t]
\centering
\includegraphics[width=\textwidth]{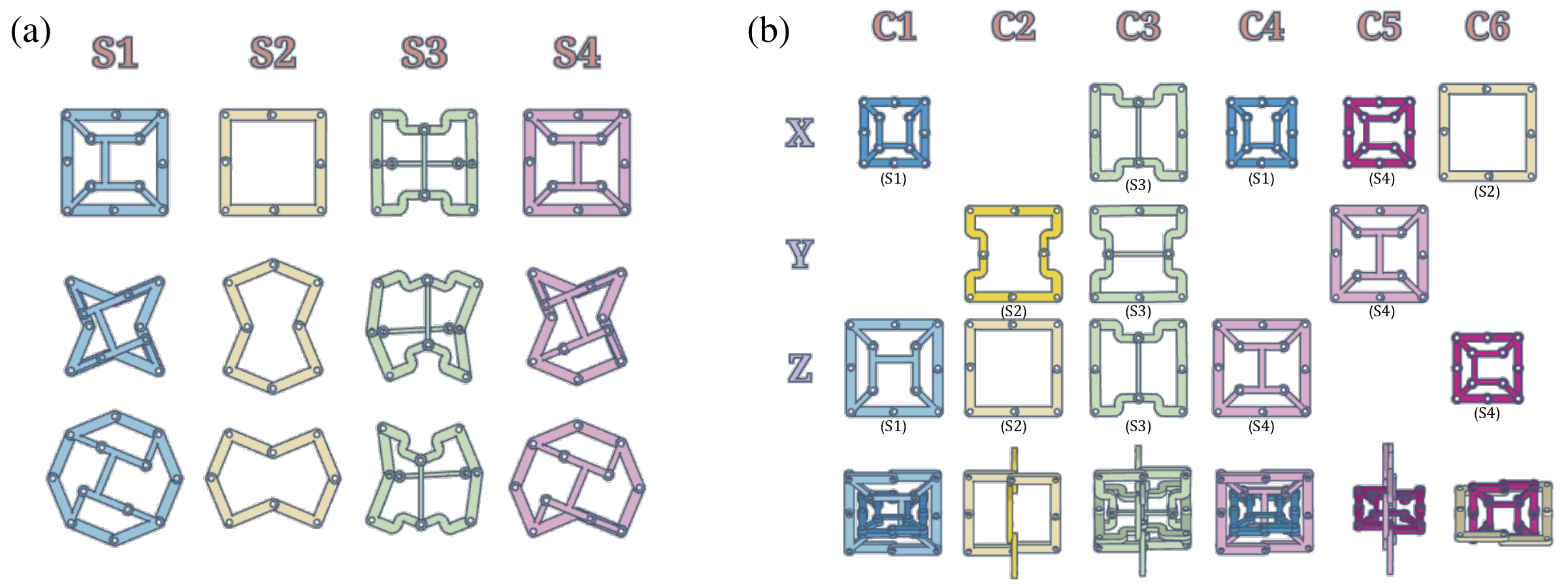}
\caption{\textbf{Design of cubic blocks.} (a) Alternative design of the square building blocks (top) and the two polarizations of each block's floppy modes (middle and bottom). (b) All cubic blocks constructed by connecting the square blocks in orthogonal planes.}
\label{fig:sup_bb}
\end{figure*}

As explained in the main text, the design of the 3D blocks is based on the fact that in each of the three planes comprising any of the cubic blocks, the behavior corresponds to one of the square blocks. We use square blocks that are slightly different than those shown in Fig ~\ref{fig:square_blocks}; We removed unnecessary beams so that we could fit a smaller square inside a bigger one in two orthogonal planes. Instead of having full triangles connecting the edges to the central square, we made the corners rigid, added a diagonal beam, and removed one of the four sides of the central square, see Fig~\ref{fig:sup_bb}a. Each of the cubes is made of two or three orthogonal square blocks, as shown in Fig ~\ref{fig:sup_bb}b. Note that square Block~S3 has to have two beams across to maintain a single floppy mode. However, when constructing the cubic Block~C3 we only need one in each plane. To connect the 3D-printed parts within each square we used LEGO\textsuperscript{\tiny\textregistered} Technic Pin without Friction Ridges 3673. To connect the two squares we used two of LEGO\textsuperscript{\tiny\textregistered} Technic Axle and Pin Connector Perpendicular 6536 and a LEGO\textsuperscript{\tiny\textregistered} Technic Axle 4L with Stop 87083. As for C3 we connected the edges with LEGO\textsuperscript{\tiny\textregistered} Technic Axle 2L with Pin without Friction Ridges 65249 and one LEGO\textsuperscript{\tiny\textregistered} Technic Axle and Pin Connector Perpendicular 6536. 

\section{C5 boundary texture parity constraint} 
\label{sec:C5_p}

\rs{There is a parity constraint on the numbers of in and out pixels on the boundary of a compatible metamaterial made of Block~C5. Let us consider a lattice of $L_x \times L_y \times L_z$ blocks, whose boundary consists of $2 (L_x L_y+L_y L_z+L_z L_x)$ pixels. If, in a certain fixed polarization, we denote by $B_i$ the number of in pixels along the boundary, then the number of out pixels will be $B_o=2(L_x L_y+L_y L_z+L_z L_x)-B_i$. If $N_i$ blocks deform with $4$ faces in and $2$ faces out, then $N_o=L_x L_y L_z -N_i$ blocks deform with $4$ faces out and $2$ faces in. Now, there are $I = L_x L_y (L_z-1) + L_y L_z (L_x-1)+ L_z L_x (L_y-1) = 3 L_x L_y L_z - L_x L_y - L_y L_z - L_z L_x$ internal faces within the bulk of the lattice. Each such internal face contributes one face in to one of the blocks sharing that face, as well as one face out of the other block, so this is the number of internal ins and also of internal outs. The number of in pixels along the boundary is thus $B_i = 4 N_i + 2 N_o  - I$, where we summed the 4 or 2 ins that the blocks in the two different states contribute, and subtracted the ins at the internal faces. After substitution, this may be written as $B_i = 2 N_i +2 L_x L_y L_z - I$. Therefore $B_i$ and $I$ share the same parity, which is even unless exactly one of the three lengths $L_x$, $L_y$, and $L_z$ is even. Since $B_i+B_o$ is even, $B_o$ has the same parity as $B_i$.}

\section{C5 texture design protocol}
\label{sec:c5 protocall}

\begin{figure*}[t!]
\centering
\includegraphics[width=\textwidth]{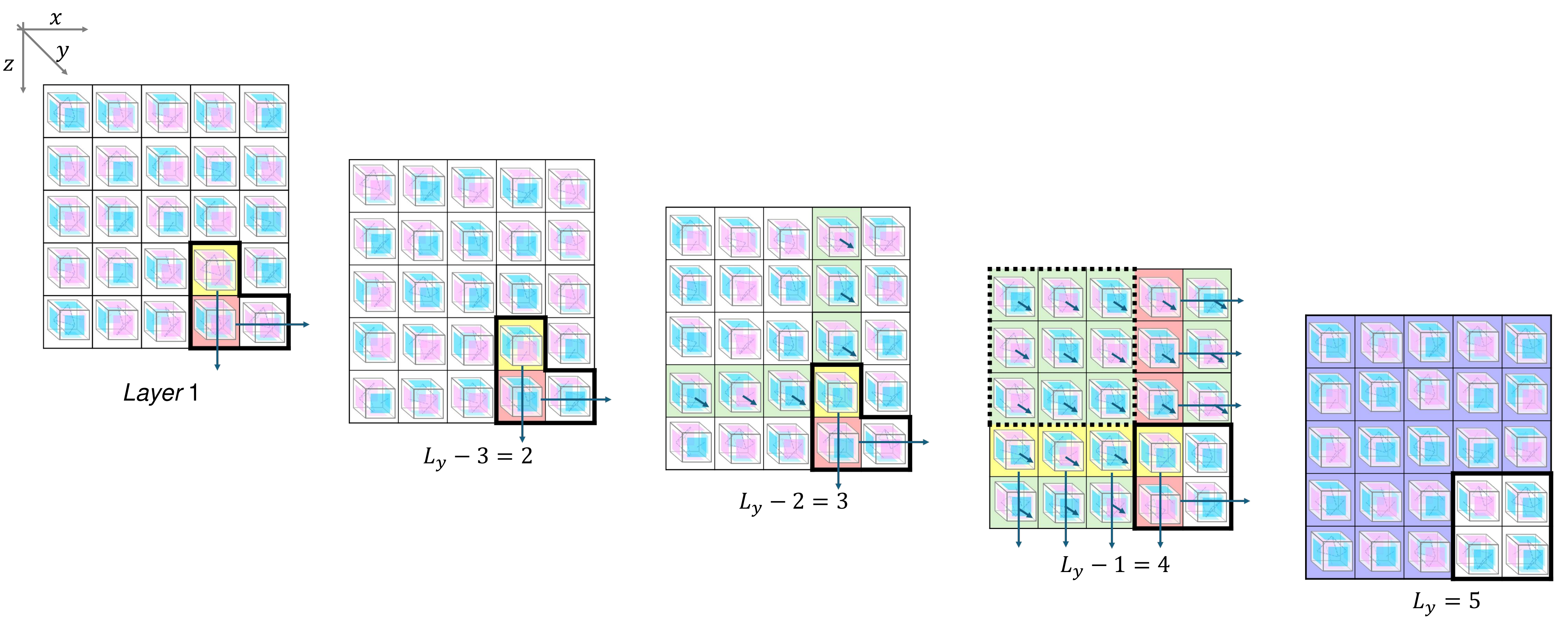}
\caption{\textbf{Block~C5 texture design.} The five layers of the lattice. In each layer we mark the cubes that need special attention when oriented. The yellow-shaded, green-shaded, and peach-shaded cells help the boundary in the $x$, $y$ and $z$ directions, respectively. In layer 5, we first orient the blocks in the bold frame, and then resolving the purple-shaded area is exactly the texture design problem with Block~S4 in the square lattice. }
\label{fig:c5_text}
\end{figure*}

We will be aided by the following basic fact about Block~C5: If the behaviors of three faces near a corner, plus one more face, are prescribed, then there is always a way to rotate the block so that its soft mode matches those four prescriptions.

We can scan the lattice layer by layer, say in the $y$ direction, from back to front. For all layers, except for the last layer, we scan row by row from top to bottom in the $z$ direction, and column by column from left to right in the $x$ direction. Each time we reach a new block, it has three adjacent faces (top, left, and back) that are set, except for blocks along the right and bottom sides of the layer that have four adjacent specified faces. Both cases can always be satisfied. The last block at the corner of the layer will have five adjacent specified faces, which include two pairs of opposite faces. To satisfy this, we make sure that one pair deforms in the same direction, specifically that the left and right faces of the corner block move in the same direction. To do so, we add some constraints on the three blocks in the corner of the layer, at positions $(L_x-1, L_z-1)$, $(L_x-1, L_z)$, and $(L_x, L_z)$ (marked in a bold frame in Fig.~\ref{fig:c5_text}) in addition to the ones they have to satisfy from the previous blocks and from the boundary. First, we set the block at position $(L_x-1, L_z-1)$, marked in yellow, in such a way that its bottom face moves the same way as the boundary texture at the bottom of the block $(L_x-1, L_z)$. Now we can add the constraint on the block at position $(L_x-1, L_z)$, marked in peach, on its right side to match the boundary pixel on the right of the block in the corner $(L_x, L_z)$. Finally, we set the block $(L_x, L_z)$, which is also possible because our procedure assures that its left and right faces move in the same direction.

We employ this procedure for all layers up to and including $L_y-3$. In layer $L_y-2$ we have to start ``helping'' the final layer that has to satisfy the boundary texture also at the front face of the lattice. We add a condition to the blocks in row $L_z-1$ and in column $L_x-1$ (marked in green in Fig.\ref{fig:c5_text}) that their front face moves in the same way as the boundary texture at the front of the equivalent blocks in layer $L_y$. All the other blocks in layer $L_y-2$ can be assigned as described above, including the three in the corner (bold frame) that are dealt with in the same way as in the previous layers. 

Now we orient the blocks in layer $L_y-1$, where our goal is to have as many as possible of the blocks' front faces ``help'' the final layer's texture. First we add the constraint to the blocks in the dotted bold region (namely between $(1,1)$ and $(L_x-2,L_y-2)$), that their front face move in the same direction as the boundary texture at the front of the equivalent blocks in layer $L_y$. This is possible because this is the fourth constraint, on adjacent sides, for these blocks. Blocks in row $L_z-1$ and column $L_x-1$ can easily have the same texture on the front, and because we prepared this in the previous layer, now we can add another condition for them. Namely, the blocks in row $L_z-1$, marked in yellow, should have their bottom faces move in the same way as the boundary texture at the bottom of the blocks in row $L_z$. Similarly, the blocks in column $L_x-1$, marked in peach, should have their right faces move in the same way as the boundary texture at the right of the blocks in row $L_x$. Now we can orient the rest of the blocks marked in green (most of row $L_z$ and column $L_x$) so that they satisfy their boundary condition as well as an added condition of having their fronts move in the same way as the outer boundary of layer $L_y$. The four blocks in the corner (in the bold frame) cannot ``help'' the final layer and will have the same orientation protocol as the previous layers, i.e., the bottom of block ($L_x-1, L_z-1$) satisfies the boundary condition on the bottom, the right side of block ($L_x-1, L_z$) satisfies the boundary condition on the right and blocks ($L_x, L_z-1$) and ($L_x, L_z$) will satisfy the neighboring blocks and their own boundary conditions.

Finally, when we get to layer $L_y$, we scan in the opposite direction. We start from the bottom right corner and first orient the four blocks in the bold frame. This is easy because in each case we only have to satisfy four sides. We already set layer $L_y-1$ so that all the blocks in the purple region will have front and back sides that move in the same direction. Block~C5 is made up of square Blocks~S3 and~S4. Once the block has one pair of opposite sides moving in the same direction, we are left with only Block~S4 to orient in the plane. Therefore, the problem at hand is the same as the one in Section~\ref{ssec:square}, of designing a boundary texture for Block~S4 in the region marked in purple. We proved there that we can do so up to one pixel, which is the same as the pixel that cannot be specified because of our parity argument for Block~C5. 

\bibliography{holography}

\end{document}